\documentclass[reprint, amsmath, amssymb, aps]{revtex4-2}

\usepackage{dcolumn}
\usepackage{bm}

\usepackage[T1]{fontenc}
\usepackage{indentfirst}
\usepackage{xcolor}
\usepackage{graphicx}
\usepackage{amssymb}
\usepackage{latexsym}
\usepackage{mathtools}
\usepackage{bbold}
\usepackage{accents}
\usepackage{hyperref} 
\usepackage[utf8]{inputenc} 
\usepackage{multirow}
\usepackage{palatino} 
\usepackage{lmodern}
\usepackage{soul}
\usepackage{array}
    \newcolumntype{P}[1]{>{\centering\arraybackslash}p{#1}}

\usepackage{adjustbox}

\newcommand\ps[1]{\mathbf{#1}^s_\mathbf{p}}

\begin{document}

\title{Non-parametric maximum likelihood component separation for CMB polarization data.}

\author{Clément Leloup$^{1,2,3}$}
 \email{clement.leloup@ipmu.jp}
\author{Josquin Errard$^{1}$}
\author{Radek Stompor$^{4,1}$}
\affiliation{$^{1}$ Université Paris-Cité, CNRS, Astroparticule et Cosmologie, F-75013 Paris, France}
\affiliation{$^{2}$ Kavli Institute for the Physics and Mathematics of the Universe (Kavli IPMU, WPI), UTIAS, \\
The University of Tokyo, Kashiwa, Chiba 277-8583, Japan}
\affiliation{$^{3}$ Center for Data-Driven Discovery, Kavli IPMU (WPI), UTIAS, The University of Tokyo, Kashiwa, Chiba 277-8583, Japan}
\affiliation{$^{4}$ CNRS-UCB International Research Laboratory, Centre Pierre Bin\'{e}truy, IRL2007, CPB-IN2P3, Berkeley, US}






\begin{abstract}
        Mitigation of the impact of foreground  contributions to measurements of Cosmic Microwave Background (CMB) polarization is a crucial step in modern CMB data analysis and is of particular importance for a detection of large-scale CMB $B$ modes. A large variety of techniques, based on different assumptions and aiming at either a full component separation or merely cleaning the foreground signals from the CMB maps, have been described in the literature.
        
        In this work, we consider this problem 
        within a unified framework based on the maximum likelihood principle, under the assumption that the signal at each frequency can be represented as a linear mixture of sky templates. We discuss the impact of various additional assumptions on the final outcome of the procedure. We find that the component separation problem can be fully solved in two specific situations: when we either know the frequency scaling of all the components or can correctly model them with a limited number of unknown parameters, as is the case in the parametric component separation techniques; or when we either know the statistical properties of all the components, the foregrounds and CMB, or can correctly model them with a limited number of parameters, as for instance in SMICA-like approaches. However, we also show that much less stringent assumptions are sufficient if we only aim at recovering the cleaned CMB signal.
        In particular, we discuss a ``minimally informed'' non-parametric method based on maximum likelihood. The method only assumes that the component properties are independent on the sky direction in some, possibly limited, region of the sky and that the CMB covariance is known up to some limited number of parameters. We apply this method to recover the CMB $B$ modes polarization signal in the context of forthcoming CMB experiments 
        and compare its performance with that of the standard parametric maximum likelihood approach. We show that the minimally informed method performs as well as the parametric one when the parametric model is correct, and supersedes it when the it is incorrect.
\end{abstract}

\maketitle

\def\matriximg{%
  \begin{matrix}
    \\
    \quad \cdots \quad \\
    \\
   \end{matrix}
}%

\section{Introduction}
\label{section:Introduction}

Cosmic Microwave Background (CMB) $B$-mode polarization offers some of the most exciting opportunities for the next stage of observational and experimental effort in CMB physics. Indeed, the measurement of large-scale CMB $B$ modes, sourced by primordial Gravitational Waves (GW), constitutes the main goal of current, (e.g. BICEP~\cite{BICEP:2021xfz}, Simons Array~\cite{Groh:2021ntp}), forthcoming (e.g. Simons Observatory~\cite{Ade:2018sbj}), and planned future CMB experiments (e.g. CMB-S4~\cite{Abazajian:2016yjj}, LiteBIRD~\cite{Litebird-2019}). The measurements of CMB $B$ modes could open up a window, as direct as likely ever possible, onto the physics of the very early Universe. This would give us unique insights on the physical laws governing at the highest energies reached during the inflation epoch. However, such outstanding results seem to be matched by difficulties, which need to be overcome in order to deliver a reliable detection and precise characterization of the primordial $B$-mode signal. The obstacles are of fundamental and instrumental origins, and many stem from the fact that the $B$-mode amplitudes are expected to be several orders of magnitude below the $B$-mode signal generated by the Galactic foregrounds~\cite{Krachmalnicoff:2015xkg}.

Polarized Galactic foregrounds are dominated by dust and synchrotron emissions, which are still only poorly known with the best constraints due to Planck~\cite{Planck:2018yye}. The limited knowledge of Spectral Energy Distributions (SEDs) of these signals leaves the possibility of a rather complex sky at the microwave band, even at high Galactic latitudes~\cite{PanEx}. This is in addition to the fact that the emission laws for synchrotron and dust may vary across the sky as suggested by recent work~\cite{Planck:2015mvg,Pelgrims:2021gqi,Miville-Deschenes:2008lza}. In the absence of reliable and physically motivated models of foreground emissions, 
statistical methods of component separation 
flexible enough to allow for all, or a sufficiently broad class of, possible sky 
models are of special interest, 
in particular in the quest of reaching the scientific target of $r \leq 0.001$ as defined for the future experimental efforts.

To tackle this crucial challenge, many different approaches of foreground removal have been developed, following various principles and assumptions. Among others, the most extensively studied are based on the empirical modeling of the foreground SEDs in parametric methods \cite{Eriksen:2005dr,Stompor:2008sf,delaHoz:2020ggq}, on statistical properties of the foreground in the Internal Linear Combination (ILC) \cite{Eriksen:2004jg,Vio:2008us,Basak:2011yt} and Independent Component Analysis (ICA) \cite{Cardoso:2008qt} techniques, and on the definition of internal or external foreground templates in template fitting methods \cite{Efstathiou:2009MNRAS.397.1355E,Katayama:2011ApJ...737...78K}. Many of these methods have been implemented as software packages, e.g.~\cite{Commander,Galloway:2022zoo,FGBuster,SMICA}, and applied extensively to real and simulated CMB data sets.

In this paper we focus on component separation techniques based on the maximum likelihood principle and discuss the role and impact of the assumptions they rely on. We do so in the context of a unified framework that we develop in Section~\ref{section:General formalism}. Then, we capitalize on the obtained insights and propose a new maximum likelihood-based, non-parametric approach to foreground cleaning with a minimal set of assumptions on foreground components. In Section~\ref{section:Implementation of the spectral likelihood}, we detail a simple implementation of the proposed method in harmonic domain, that we use on study cases relevant for future experiments in Section~\ref{section:Performances of the method}.

\section{General formalism}
\label{section:General formalism}

\subsection{Data model}
\label{subsection:Data model}

We assume that our input data set is composed of data collected in multiple frequency bands and that the signal detected in each of these is a linear combination of signals of different physical origins, referred to as components, with coefficients depending on the frequency band. These component amplitudes are collected into a single vector denoted hereafter as $\bold{s}$, and the coefficients into a single matrix from component space to frequency space, denoted as $\bold{A}$ and referred to as the mixing matrix. The component signals can be expressed in different domains depending on the context, be it pixel-domain, i.e. when the data are sky maps, or their spherical harmonic representation. For concreteness, we work in this section in the pixel domain but the generalization is straightforward. Denoting by $\bold{d}$ the data vector combining all measured frequency signals and by $\bold{n}$ the noise, we can write down mathematically our data model as,
\begin{equation}
    \bold{d} = \bold{A}\,\mathbf{s} + \bold{n}. \label{eq:data model}
\end{equation}
In the component separation problem, the mixing matrix elements are in general unknown.
We will make, however, some additional assumptions in the following. In particular, we will assume that the mixing matrix does not combine information coming from different sky pixels (or harmonic modes respectively) nor does it combine different Stokes parameters in the case of polarization-sensitive experiments. These assumptions may not be always fulfilled, in particular in the presence of instrumental effects, but are sufficiently general for this work. Consequently, the full mixing matrix $\mathbf{A}$ has a hierarchical block-structure with blocks corresponding to Stokes parameters, denoted $\bold{A}^{s}$ , which can be further broken down into pixel specific sub-blocks $\ps{A}$. We will also use a subscript $\bold{p}$ to denote pixel-specific objects, e.g. $\bold{A}_\mathbf{p}$ will denote a pixel-specific block of  the matrix $\mathbf{A}$ which in general will combine information relevant to all Stokes parameters. The most basic blocks of the mixing matrix $\ps{A}$, are furthermore assumed to have a full-column rank. If this is not the case, this would mean that some of the allowed-for components are not independent for a given set of observational frequency bands and should be therefore merged together, thus reducing the number of columns of the mixing matrix and restoring independence of its columns. If the number of components is assumed to be known, yet another important albeit usually implicit assumption, then the number of frequency channels has to be at least as large as the number of components. In the following we will always assume that this is indeed the case and that the mixing matrix columns are independent for the assumed number of components and the available frequency bands.

We will denote the number of sky components as $n_{c}$ and the number of frequency channels as $n_{\nu}$. Consequently, for each considered Stokes parameter, $s$, and for each sky pixel, $\bold{p}$, the corresponding data vector $\ps{d}$ and the noise vector $\ps{n}$ both have $n_{\nu}$ elements. The mixing matrix $\ps{A}$ is a dense matrix with $n_{\nu}$ rows and $n_{c}$ columns, and the component vector $\ps{s}$ has a number of entries given by $n_{c}$. By assumption, we have $n_{\nu} \geq n_{c}$. The mixing matrix $\ps{A}$ will be generally unknown and can have as many as $n_{\nu}\,n_{c}$ unknown elements. Together with $n_{c}$ unknown component amplitudes, that leads to as many as $n_{c} \, \left( n_{\nu} + 1 \right)$ unknown parameters which we would like to estimate from the data. This number always exceeds the number of observations for any single pixel. Indeed, for a single pixel we can set at most $n_\nu-n_c$ constraints on the mixing matrix elements and $n_c$ constraints on the component amplitudes. However, if the component mixing can be assumed to be the same for a number of sky pixels then the number of data points increases and can surpass the number of unknown mixing matrix elements and component amplitudes. As we discuss below this allows to relax the single pixel limit of $n_\nu-n_c$ on the number of constraints which can be set on the mixing matrix elements under some rather generic assumptions. This, however, still does not allow to estimate the mixing matrix completely without further assumptions. In the following we discuss what limit can be set given different additional assumptions and in turn, what assumptions are needed to solve the component separation problem. To do so, we cast the problem in the framework of maximum likelihood formalism. However, a generalization to the cases involving multiple mixing matrices corresponding to different disjoint subsets of all the pixels is straightforward.


\subsection{Maximum Likelihood approach}
\label{subsection:Maximum Likelihood approach}

We assume that the noise in input single frequency maps is drawn from a multivariate Gaussian distribution with known covariance $\mathbf{N}$ and zero mean. We can therefore define the log-likelihood to minimize to be,
\begin{equation}
    \mathcal{S} \left( \bold{n} \right) \equiv -2\ \mathrm{ln} \ \mathcal{L} \left( \bold{n} \right) = \bold{n}^{T}\bold{N}^{-1}\bold{n}.
    \label{eqn:likeNoise}
\end{equation}
We can rewrite this likelihood as a function of the unknown parameters, assuming the data model in Eq.~\eqref{eq:data model},
\begin{equation}
	 \mathcal{S} \left( \bold{A}, \bold{s} \right) = \left( \bold{d} - \bold{A} \bold{s} \right)^{T} \bold{N}^{-1} \left( \bold{d} - \bold{A} \bold{s} \right).
	 \label{eq:likelihood}
\end{equation}
Because it only depends on the product $\mathbf{A}\,\mathbf{s}$, this likelihood is invariant under a simultaneous transformation of the mixing matrix and the sky amplitudes that leaves their product invariant,
\begin{equation}
    \bold{A} \rightarrow \bold{B} = \mathbf{A}\mathbf{M}^{-1} \qquad \text{and} \qquad \mathbf{s} \rightarrow \mathbf{s'} = \mathbf{M}\,\mathbf{s},
    \label{eq:invarianceGeneral}
\end{equation}
where $\bold{M}$ is an arbitrary invertible matrix of dimensions $n_{c}\times n_c$ that we call in the following the transformation matrix. This transformation can be interpreted as a change in the definition of the components and thus of their frequency scaling in such a way that they reproduce the same combined signal emissions in all observed frequencies. The likelihood treats every such linear combination of the physical components as equivalent in the absence of any additional information. Therefore, in general multiple solutions to the component separation problem exist and more assumptions are needed to lift this degeneracy. These can concern either the mixing matrix itself, e.g. frequency scaling of the sky components, or the components themselves, e.g. their morphology or statistical properties.

We can gain more insight into this problem by considering the so-called spectral likelihood as introduced in~\cite{Stompor:2008sf}. It can be obtained from Eq.~\eqref{eq:likelihood} via maximization with respect to the component amplitudes $\mathbf{s}$. Indeed, for any choice of the mixing matrix elements $\mathbf{A}$, the maximum likelihood solution of Eq.~\eqref{eq:likelihood} for the component amplitudes is given by the generalized least-square estimator,
\begin{equation}
	\bold{\bar{s}} = \left( \bold{A}^{T}\bold{N}^{-1}\bold{A} \right)^{-1} \bold{A}^{T}\bold{N}^{-1} \bold{d}.
	\label{eq:skyEstim}
\end{equation}
Inserting it back in Eq.~\eqref{eq:likelihood} yields the spectral likelihood,
\begin{eqnarray}
	\mathcal{S}_{\mathrm{spec}} \left( \mathbf{A} \right) & \equiv
    & \mathcal{S} \left( \mathbf{A}, \bold{\bar{s}} \right) \nonumber \\
	& = & - \bold{d}^{T} \bold{N}^{-1} \bold{A} \left( \bold{A}^{T}\bold{N}^{-1}\bold{A} \right)^{-1} \bold{A}^{T}\bold{N}^{-1} \bold{d}. 
	\label{eq:genericSpectralLikelihood}
\end{eqnarray}
On maximizing the spectral likelihood with respect to some parametrization of the mixing matrix, we can derive the same maximum likelihood values as from the initial likelihood in Eq.~\eqref{eq:likelihood} \cite{Stompor:2008sf}. Moreover, as long as the transformation $\mathbf{M}$ is invertible we have,
\begin{eqnarray}
\mathcal{S}_{\mathrm{spec}} \left( \mathbf{A} \right) & = & \mathcal{S}_{\mathrm{spec}} \left( \mathbf{B} =\bold{AM}^{-1} \right).
\end{eqnarray}
Consequently, if some matrix $\mathbf{A}$ is a maximum likelihood solution of Eq.~\eqref{eq:genericSpectralLikelihood} then so is any matrix $\mathbf{B}\,=\,\mathbf{A}\,\mathbf{M}^{-1}$, for any invertible $\mathbf{M}$. Similarly, the solution for the components is not unique either, since we have from Eq.~\eqref{eq:skyEstim} that,
\begin{eqnarray}
	\bold{\bar{s}} & = & \mathbf{M}^{-1}\,\left( {\bold{B}}^{T}\bold{N}^{-1}\bold{B}\right)^{-1} {\bold{B}}^{T}\bold{N}^{-1} \bold{d},
	\label{eq:skyEstim1}\\
	& = & \mathbf{M}^{-1}\,\mathbf{\bar{s}'},
	\label{eq:skyEstim2}
\end{eqnarray}
and both $\mathbf{\bar{s}}$ and $\mathbf{\bar{s}'}$ are maximum likelihood estimates corresponding to two different choices for the mixing matrix. These indeed fulfill,
\begin{eqnarray}
\mathbf{A}\,\mathbf{\bar s} \; = \; \mathbf{B}\,\mathbf{\bar{s}'}.
\end{eqnarray}

Solving the component separation problem requires therefore additional assumptions in order to remove, or at least reduce, the freedom in choosing the matrix $\mathbf{M}$, effectively breaking the degeneracy in the mixing matrix parameters. To solve the component separation problem completely, the choice of the matrix $\mathbf{M}$ has to be limited to the unit matrix. However, if the allowed matrices $\mathbf{M}$ can be restricted to matrices of the form,
\begin{equation}
    \ps{M} = \left[ \begin{array}{c|c}
  		\rule[-1.2ex]{0pt}{0pt} 1 & \mathbf{0}_{n_c-1}^{T} \\
		\hline
		\rule{0pt}{2.5ex} \mathbf{0}_{n_c-1} & \ps{Q}
	\end{array} \right], \label{eq:MdefForCleaning}
\end{equation}
then we could estimate a unique, foreground-cleaned CMB map, to the detriment of reconstructing foreground components. This follows from Eq.~\eqref{eq:skyEstim2}, given the CMB amplitudes in $\mathbf{\bar{s}}'$ and in $\mathbf{\bar{s}}$ are bound to be the same in this case.
Here, $\mathbf{Q}$ is some invertible matrix, $\mathbf{0}_{n_c-1}$ is a vector of $n_c-1$ zeros and, for definiteness, we hereafter assume that the CMB signal is stored as the first element of the signal amplitude vector for each pixel.  

Restricting the freedom of possible matrices $\mathbf{M}$ can be done by introducing additional prior knowledge
which do not conform with the transformation in Eq.~\eqref{eq:invarianceGeneral} for any general matrix $\mathbf{M}$. In the following, we consider some cases of physically motivated priors and discuss their consequences on the solvability of the component  separation problem.

\subsubsection{Impact of prior knowledge.}
\label{subsubsection:Possible mixing matrix parametrizations}

\paragraph{No prior information.}
\ \\
This corresponds to the case of a fully general mixing matrix. 
For each pixel $\mathbf{p}$ and Stokes parameter $s$, we can represent the corresponding block of the mixing matrix $\mathbf{A}^s_\mathbf{p}$ as, 
\begin{equation}
    \ps{A} = \left[ \begin{array}{c}
    \rule[-2.5ex]{0pt}{25pt} \bold{M}_\mathbf{1,p}^s \\ \hline
    \rule[-2.5ex]{0pt}{25pt} \mathbf{A}_\mathbf{p}^{\prime s} \\
    \end{array} \right] = \left[ \begin{array}{c}
    \rule[-2.5ex]{0pt}{25pt} \mathbb{1}_{n_{c}} \\ \hline
    \rule[-2.5ex]{0pt}{25pt} \mathbf{A}_\mathbf{p}^{\prime s} \left( \bold{M}_\mathbf{1,p}^s \right)^{-1} \\
    \end{array} \right] \bold{M}_\mathbf{1,p}^s \equiv \bold{B}_\mathbf{1,p}^s\mathbf{M}_\mathbf{1,p}^s,\label{eqn:genAbreakdown}
\end{equation}
where $\bold{M}^s_\bold{1}$ is an $n_{c}\times n_c$ invertible matrix defined up to a rearrangement of the rows, i.e. ordering of the frequency channels.
We note that, fixing the upper block of the mixing matrix minimizes the number of free elements of the redefined mixing matrix $\mathbf{B}$ in the absence of prior information,
reducing it from the initial number of $n_\nu\,n_c$ unknown elements in the mixing matrix $\mathbf{A}^s$ down to $(n_\nu-n_c)\,n_c$ elements in $\mathbf{B}^s_\mathbf{1}$.

Using the spectral likelihood, Eq.~\eqref{eq:genericSpectralLikelihood}, we can set at most constraints on all free elements of the matrix $\mathbf{B}_\mathbf{1}$, i.e. at most $n_{s}\,n_{c}\, \left( n_{\nu} - n_{c} \right)$ constraints in total
per pixel. As we assume that $n_{c} \leq n_{\nu}$, these numbers of parameters exceed the number of single pixel observations given by $n_s\,n_\nu$, which limits the number of possible mixing matrix constraints to $n_s\,(n_\nu - n_c)$. Consequently, we can not ever determine the matrix $\mathbf{B}_\mathbf{1}$, if only single pixel data are available. 
For completeness, we note that when the number of frequency channels is exactly equal to the number of components, no constraints can be set on the elements $\mathbf{B}_\mathbf{1}$, however, the matrix then does not contain any free elements either as $\mathbf{B}_\mathbf{1} = \mathbb{1}$ and $\mathbf{A} = \mathbf{M}_\mathbf{1}$. In this case, the components are identified with the $n_c$ available frequency maps.

\begin{figure}[!htb]
\begin{center}
\includegraphics[width=\columnwidth]{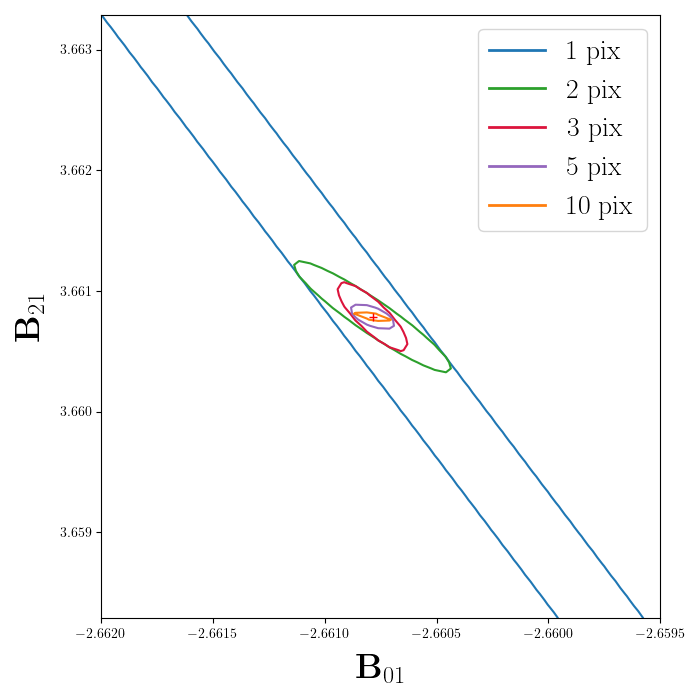}
\caption{One sigma contours of the free mixing matrix elements of the matrix $\bold{B}$ in the fully general mixing matrix case for various values of $n_{\rm pix}$ in a fictitious noiseless case where the sky is composed of CMB and one dust component only, and observations are made in three frequency channels. The red cross corresponds to the values of the parameters used to produce input maps. We can set only one constraint on the mixing matrix elements having only single pixel data (Blue lines). However, adding data of more pixels breaks the degeneracy and progressively improves the constraints.}
\label{fig:evolution with pixel number}
\end{center}
\end{figure}

In some circumstances, we can however put meaningful constraints on all the elements of $\mathbf{B}_\mathbf{1}$. This is the case when the same matrix can be applied to multiple pixels. Indeed, the total number of parameters we want to derive, including the mixing matrix elements and the component amplitudes, is given by $n_{c} \left( n_{\nu} + n_\mathrm{pix} - n_{c} \right)$ parameters for total intensity and $n_{c} \left( n_{\nu} + 2\,n_\mathrm{pix} - n_{c} \right)$ for polarization, as in this latter case we usually assume the same mixing matrix for both Stokes parameters, $Q$ and $U$. This can be lower than the total number of measurements $n_\mathrm{pix}\,n_s\,n_\nu$ if the number of pixels in which the mixing elements are constant is sufficiently large, i.e. $n_\mathrm{pix} \ge n_c$ for total intensity and $n_\mathrm{pix} \ge n_c/2$ for polarization. Indeed, these upper limits can be saturated as shown in FIG.~\ref{fig:evolution with pixel number}. For this to happen some additional assumptions have to be fulfilled. For one, the component signals need to be different in at least $n_\mathrm{pix} \ge n_c$ for total intensity and $n_\mathrm{pix} \ge n_c/2$ for polarization. In other words, only the pixels separated by more than a single correlation length of the assumed sky components effectively contribute to setting the constraints in access of the single pixel limit of $n_\nu-n_c$.
Once the saturation is reached, increasing the number of pixels in the region where the mixing elements are constant will help reducing the statistical uncertainties of recoverable parameters but can not constrain any additional parameters.

We note that an interpretation of these observations is rather intuitive. Having made no assumptions about the sought-after sky components, any linear combination of them is as good as any other. Consequently, we can not do any better than to select $n_c$ frequency channels and assume that the signal contained in these defines the components. This corresponds to setting all elements of the rows of these channels in the mixing matrix to 0 except one which is set to 1. This non-zero element appears at the position corresponding to the component number this channel is set to define. All the other channels are then arbitrary linear combinations of the selected ones with unknown coefficients. The mixing matrix corresponding to such a case is given, up to channels reordering, by the matrix $\mathbf{B}^s_\mathbf{1}$ defined in Eq.~\eqref{eqn:genAbreakdown}. Following the discussion above we can conclude that, if it can be assumed that the mixing is the same for at least $n_c$ sky pixels for which the component amplitudes are different, we can set constraints on all the unknown mixing coefficients. And these coefficients define the frequency dependence of the sky components as defined above. 

While mathematically consistent this is not satisfactory from the physics point of view in most practical applications where the components are associated with actual physical processes. In order to go any further we need to introduce more assumptions based on physical insights, which will allow to identify the components. We discuss some options in the following. We emphasize that, in any case, the solutions obtained here will be only as good as the physics-driven assumptions and different ways of identifying the component may lead to different results. \\

\paragraph{Prior on CMB frequency scaling.}
\ \\
We consider the case  where the frequency scaling of the CMB signal is assumed to be perfectly known. This assumption is very well motivated physically as both theory and observations support the fact that the frequency scaling of CMB is very close to that of a black body~\cite{Fixsen:1996nj}. It formally means that the column of the mixing matrix corresponding to the CMB component is known and fixed, and only the transformation matrices $\mathbf{M}$ which leave it invariant are admissible. For concreteness and without loss of generality, we assume that all the data is expressed in CMB units and that the CMB contributes to all frequency channels. This fixes all the elements of the CMB columns of the mixing matrix to 1, i.e.,
\begin{equation}
    \bold{A}_\mathbf{p}^s = \left[ \begin{array}{c|c}
    1 & \\
    \vdots & \mathbf{A}_\mathbf{f, p}^s \\
    1 & 
    \end{array} \right] \equiv \left[ \begin{array}{c|c}
     & \\
    \bold{A_{c}} & \mathbf{A}_\mathbf{f, p}^s \\
     & 
    \end{array} \right], \label{eq:decomposition mixing}
\end{equation}
where $\mathbf{A}_\mathbf{c}$ and $\mathbf{A}_\mathbf{f}$ describes respectively the CMB and the foreground component mixing. This shape of the mixing matrix restricts the transformation matrix $\mathbf{M}$ to be of the following form,
\begin{equation}
	\mathbf{M}^s_\mathbf{2, p} = \left[ \begin{array}{c|c}
	1 & \bold{m}^{T} \\ \hline
	& \\
	\mathbf{0}_{n_c-1} & \mathbf{\hat{M}}^s_\mathbf{2, p} \\
	& 
	\end{array}	 \right],
\end{equation}
where $\mathbf{\hat{M}}^s_\mathbf{2, p}$ is an arbitrary $ (n_{c} - 1) \times (n_c - 1)$ invertible matrix and $\mathbf{m}$ is a $n_c-1$ vector. We can choose the transformation matrix $\mathbf{M}^s_\mathbf{2}$, so that the redefined mixing matrix is given by,
\begin{equation}
    \mathbf{B}_\mathbf{2, p}^s \; = \; \mathbf{A}_\mathbf{p}^s\;{\mathbf{M}^s_{\mathbf{2}, \mathbf{p}}}^{-1} \; = \; \left[ \begin{array}{c|c}
		 & \rule[-1.2ex]{0pt}{0pt} \mathbf{0}_{n_c-1}^{T} \\ \cline{2-2}
		 & \\
		 & \boldsymbol{\mathbb{1}}_{{n_{c}-1}} \\
		 \bold{A_{c}} & \\ \cline{2-2}
		 & \\
		 & \mathbf{\hat B}^s_\mathbf{2, p} \\
		 & \\
	\end{array}	\right]. \label{eqn:aPrimeFixedBlock}
\end{equation}
This can be obtained by taking $\mathbf{\hat M}_\mathbf{2, p}^s$ to be,
\begin{eqnarray}
\mathbf{\hat M}_\mathbf{2, p}^s & \equiv &  \mathbf{A}^s_{\mathbf{p}\; [1:n_c-1, 1:n_c-1]} \; - \;
\left[
\begin{array}{@{}c@{}}
1\\
\vdots\\
1
\end{array}
\right]
\;
\mathbf{m}^{T};
\label{eqn:blocksDef}
\end{eqnarray}
where the first term on the right hand side stands for a sub-block of the matrix $\mathbf{A}$ with ranges as defined by the indices (the indices of $\mathbf{A}$ range in $\left[ 0,n_{\nu}-1 \right]$ for the rows and in $\left[ 0,n_{c}-1 \right]$ for the columns) and $\mathbf{m}^{T} = \mathbf{A}_{\mathbf{p}\; [0, 1:n_c-1]}$, collecting the non-CMB elements from the first row of $\mathbf{A}_\mathbf{p}$.

Given that the resulting $\mathbf{M}_\mathbf{2}^s$ is neither a unit matrix nor a matrix of the form as in Eq.~\eqref{eq:MdefForCleaning}. Therefore, only fixing the CMB frequency dependence is not enough to solve the component separation problem, or even merely to clean the CMB of foregrounds. Even though in practice some methods, such as ILC, only make this assumption and show satisfactory results \cite{Planck:2018yye}, these are fundamentally imperfect in theory and must demonstrate their performance on a case by case basis. The only advantage brought by the knowledge of the CMB scaling over the fully general case is that the CMB is now confined in a single redefined component of $\mathbf{s'}$ given by the frequency channel corresponding to the first row of the matrix $\mathbf{B}^s_\mathbf{2}$. All other recovered sky components in $\mathbf{s'}$ will then contain only foregrounds. This can be seen from Eq.~\eqref{eq:skyEstim2} given the form of $\mathbf{M}_\mathbf{2}$. This is again quite intuitive as the knowledge of the CMB scaling allows us to rescale all the channels in a way that the CMB amplitudes are the same in all of them, i.e. effectively expressing them in CMB units. Then, by subtracting the channel corresponding to the first row of the mixing matrix from all the others, we can remove the CMB part from them. The CMB signal is then confined to the first channel while the remaining $n_\nu-1$ CMB-free channels can be thought of as a new, reduced data set containing only $n_c-1$ foreground components for which the conclusions of the fully general mixing matrix case apply.

We note that in this case the number of constraints which can be set on the mixing matrix elements is reduced and given by $\left( n_c-1 \right) \left( n_\nu-n_c \right)$ for both total intensity and polarization, reflecting the fact that effectively we deal with only $n_c-1$ components. \\

\paragraph{CMB scaling with foreground templates.}
\ \\
Let us assume now that not only the CMB scaling is known but also we can identify $n_c-1$ channels for which the CMB contribution is negligible. 
Assuming that these channels are linearly independent, they can be used to define the foreground components. The full mixing matrix reads in such a case, up to a reordering of the frequency channels,
\begin{equation}
	\mathbf{A}^s_\mathbf{p} = \left[ \begin{array}{c|c}
	& \\
	\mathbf{0}_{n_c-1} & \mathbf{A}^s_\mathbf{t, p} \\
	& \\ \hline
	1 & \\
	\vdots & \mathbf{A}^s_\mathbf{f, p} \\
	1 & 
	\end{array}	 \right],
\end{equation}
where $\mathbf{A_t}$ defines the foreground templates and $\mathbf{A_f}$ describes the mixing of foreground components in the channels where CMB is present.

In spite of appearances, this case is equivalent to the previous case. Indeed, as we discussed there we can always take one channel and subtract, after a rescaling of all the channels to the CMB units, removing the CMB component from $n_c-1$ channels and therefore converting them into the foreground-only templates. This procedure introduces cross-channel noise correlations because the template frequency channels share the same noise contribution from the subtracted channel. As this operation is fully invertible it will not affect the conclusions of the previous case.

Similarly to the previous case, these assumptions are not sufficient to solve the component separation problem or even to recover a foreground cleaned CMB map. While this may look somewhat counter-intuitive, the reason for this is that while foreground templates define uniquely the space of all possible foregrounds, we still cannot recover unique mixing matrix coefficients needed to subtract those from the data to recover the CMB. We could instead deproject those from any input single frequency maps but this procedure, in the absence of any additional assumptions, will in general bias the recovered CMB map by removing all the CMB signal which resides in the space spanned by the templates. We discuss sufficient assumptions to correct for that later on. \\

\paragraph{Priors on frequency scaling of sky components.} 
\ \\
Let us assume that frequency scaling of some of the components is known. If so, as in the case of the known CMB frequency scaling, we can reduce the problem to that of the general mixing matrix case with fewer sky component. Indeed, we can rescale all the channels so the mixing matrix elements corresponding to one of these components are set to 1 and then remove it from the considerations thus reducing the problem. We can then continue this procedure sequentially for each of the other components with known scaling, obtaining a reduced data set with fewer channels and components. This final data set will have a fully general mixing matrix as in one of the cases discussed earlier. We note that even in the case when the scaling for all components but one is known we still can only constrain the relative ratios of the mixing matrix elements corresponding to that component and the left-over frequency bands after the subtractions, which does not allow the problem to be fully solved. Indeed, in the absence of any other type of assumptions, knowledge of the scaling laws for all the components is a necessary requirement to solve the component separation problem. Such a requirement is also manifestly sufficient.

Interestingly, however, we do not need to know any of these scaling laws exactly and instead can afford to have up to $n_\nu-n_c$ unknown parameters parametrizing each of them. Under some general assumptions about those, the degeneracy in Eq.~\eqref{eq:invarianceGeneral} will be broken as any transformation with $\mathbf{M}\ne \mathbb{1}$ will unavoidably affect the scaling of the components. At the same time, the number of constraints on mixing matrix elements we can set is the same as in the case of the general mixing matrix, i.e., $(n_\nu-n_c)\,n_c$, in total for each Stokes parameter or $n_\nu-n_c$ per component and each Stokes parameter. This defines the maximum number of free parameters in scaling laws for each sky component. As before, this limit can be only saturated if the same mixing is assumed for multiple pixels.

This last observation is indeed the basis of all the parametric maximum likelihood component separation approaches~\cite{Planck:2018yye,Eriksen:2005dr,Stompor:2008sf}. These methods are extensively studied in the literature and we use the implementation of~\cite{Stompor:2008sf,Stompor:2016hhw} as a reference method. Importantly, following the discussion in this section, we can conclude that we can constrain more spectral parameters overall than it is typically considered if we can assume the same scaling for multiple pixels. The standard limit of $n_\nu-n_c$, however, applies if different parameter values are allowed for each pixel. \\

\paragraph{Priors on component amplitudes.}
\ \\
So far in all the cases we have considered, we have been making assumptions merely concerning the mixing matrix. This is obviously not the only way in which the freedom in defining the components in Eq.~\eqref{eq:likelihood} can be restricted. This can also be achieved by introducing our expectations about some properties of the underlying components. Most readily, this can be done by adding a prior concerning statistical properties of the sky components. For instance, if the components are expected to have different signal RMS values, then the freedom of rotating one into another with the transformation matrix $\mathbf{M}$ will be limited, potentially allowing for solving the full component separation problem.

The statistical properties of the foregrounds are not well known and including such a prior may be only straightforward for the CMB signal as it is expected, and has been measured, to be nearly Gaussian~\cite{Planck:2019kim}. However, if no prior is applied to the foregrounds, this will still not remove all the freedom in their definition and any transformation matrix $\mathbf{M}$ of the form as in Eq.~\eqref{eq:MdefForCleaning} will still be acceptable.
Thus, while not allowing to solve the full problem, it would render possible to clean the CMB map of the foregrounds, at least in principle.

This is the option we study in the reminder of this work.

\ \\
Before concluding this section, we would like to emphasize that these various cases are merely some of the most common possible examples. 
They are not only not exhaustive but can also be combined together, leading to various forms of the mixing matrix $\bold{A}$ and of the corresponding degeneracy in its parameters determined by $\bold{M}$.

\subsection{Maximum likelihood, non-parametric foreground cleaning}

\begin{figure}[!htb]
\begin{center}
\includegraphics[width=\columnwidth]{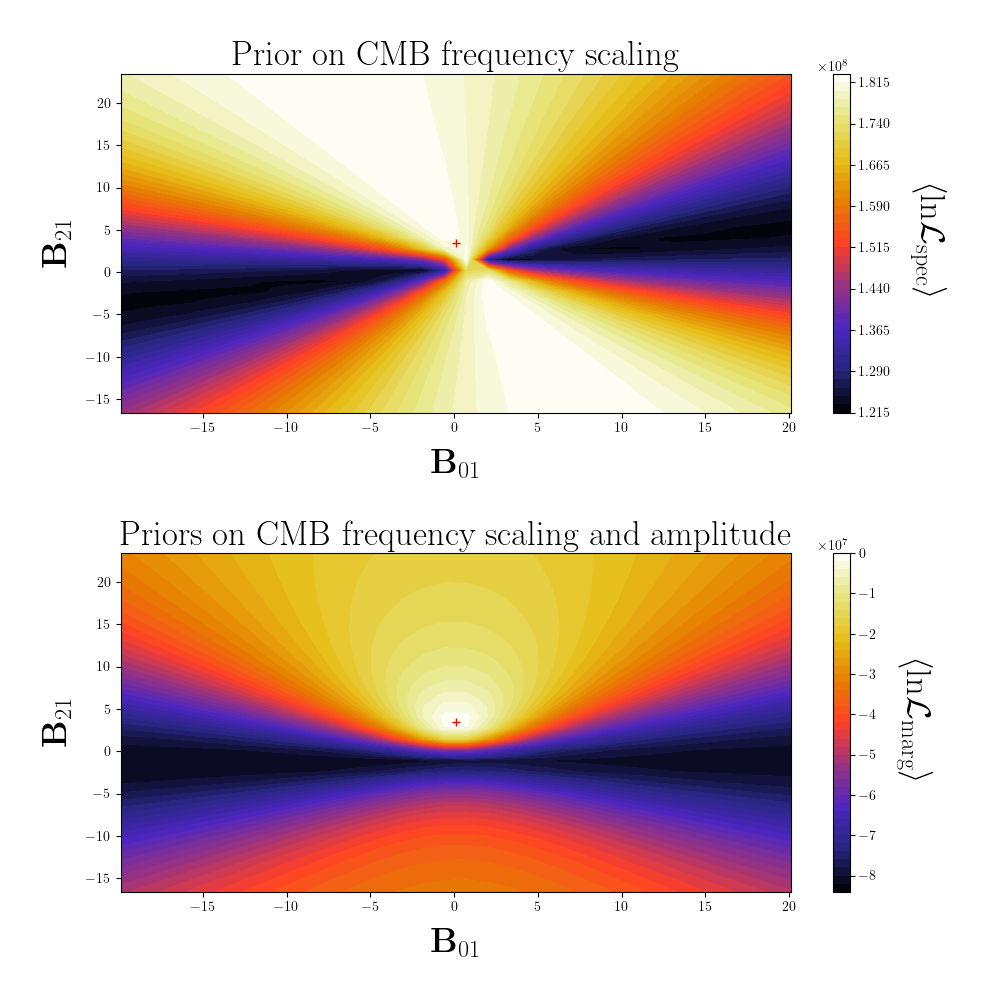}
\caption{Value of the likelihood, Eq.~\eqref{eq:genericSpectralLikelihood}, (top) and of the likelihood, Eq.~\eqref{eq:specLikeDouble Marginalized}, (bottom) when the mixing matrix is of the form of Eq.~\eqref{eq:mixing matrix form} in a fictitious noiseless case where the sky is composed of CMB and one dust component only, and observations are made in three frequency channels. The red cross corresponds to the values of the parameters used to produce input foreground maps.}
\label{fig:degeneracy}
\end{center}
\end{figure}

Following the results of the previous section, we will assume hereafter a known CMB frequency scaling and a Gaussian prior on the CMB component with covariance $\mathbf{S}_\mathbf{c}$. Introducing the prior will modify the likelihood given by Eq.~\eqref{eq:likelihood} into,
\begin{eqnarray}
	\mathcal{S} \left( \mathbf{A}, \mathbf{s}, \bold{S_{c}} \right) & = & \left( \bold{d} - \bold{A} \bold{s} \right)^{T} \bold{N}^{-1} \left( \bold{d} - \bold{A} \bold{s} \right) \nonumber \\
	&& + \ \bold{s_{c}}^{T} \bold{S_{c}}^{-1} \bold{s_{c}} + \mathrm{ln} \left| \bold{S_{c}} \right|,
	\label{eq:likelihood with prior}
\end{eqnarray}
where $\bold{s_{c}}$ stands for the CMB signal component. We see that while there is still freedom in defining the foreground components, we are not free to mix the CMB and foreground signals anymore due to the presence of the prior term. This limits the transformation matrix $\mathbf{M}$ to be of the form of Eq.~\eqref{eq:MdefForCleaning}. It is, thus, possible to find a transformation matrix $\mathbf{M}_\mathbf{3, p}^s$ of this form such that the corresponding mixing matrix $\mathbf{A}^s$ is given by,
\begin{eqnarray}
	\bold{A_{p}}^s & = & \left[ \begin{array}{c|c}
		 & \rule[-1.2ex]{0pt}{0pt} { \bold{a}^s_\mathbf{f, p}}^{T} \\ \cline{2-2}
		 & \\
		 & \bold{A}^s_\mathbf{f_{1}, p} \\
		 \bold{A_{c}} & \\ \cline{2-2}
		 & \\
		 & \bold{A}^s_\mathbf{f_{2, p}} \\
		 & \\
	\end{array}	 \right] 
      =  \left[ \begin{array}{c|c}
		 & \rule[-1.2ex]{0pt}{0pt} {\displaystyle \bold{a_{f, p}}^{T}\,[\bold{A}^s_\mathbf{f_{1}, p}]^{-1}} \\ \cline{2-2}
	
		 & \\
		 &{\displaystyle \mathbb{1}_{n_{c}-1}} \\
		 {\displaystyle \bold{A_{c}} } & \\ \cline{2-2}
		 & \\
		 & {\displaystyle \bold{A}^s_{\mathbf{f_{2}, p}}[\mathbf{A}^s_\mathbf{f_{1}, p}]^{-1}} \\
		 & \\
	\end{array}	 
     \right] \bold{M}_\mathbf{3, p}^s \nonumber \\
     & \equiv & \bold{B}_\mathbf{3, p}^s\mathbf{M}_\mathbf{3, p}^s,  
	\label{eq:mixing matrix form}
\end{eqnarray}
where the matrix $\bold{M}_\mathbf{3, p}^s$ is defined to be,
\begin{equation}
    \bold{M}_\mathbf{3}^s = \left[ \begin{array}{c|c}
  		\rule[-1.2ex]{0pt}{0pt} 1 & \bold{0}_{n_{c}-1}^{T} \\
		\hline
		\rule{0pt}{2.5ex} \bold{0}_{n_{c}-1} & \bold{A}^s_\mathbf{f_{1}, p}
	\end{array} \right],
\end{equation}
in agreement with Eq.~\eqref{eq:MdefForCleaning}. As before, we can constrain only the elements of $\bold{B}_\mathbf{3}$ but not those contained in the transformation matrix $\mathbf{M}_\mathbf{3}$. The number of free elements of $\mathbf{B}_\mathbf{3, p}^{s}$ is $\left( n_{c}-1 \right) \left( n_{\nu} + 1 - n_{c} \right)$, which is larger by $n_c-1$ than in the case of the CMB frequency scaling prior. This is illustrated in FIG.~\ref{fig:degeneracy} that shows constraints on the two mixing elements set by the spectral likelihood without prior on the CMB signal amplitude, Eq.~\eqref{eq:genericSpectralLikelihood}, and with prior on the CMB signal amplitude, Eq.~\eqref{eq:specLikeDouble Marginalized}. The mixing matrix is parametrized as in Eq.~\eqref{eq:mixing matrix form}. We see that the degeneracy between the parameters seen in the absence of prior is broken when the CMB prior is added.

These are exactly these additional constraints which permit the removal of the foreground components from the CMB signal. However, as the transformation matrix still retains degrees of freedom, not all parameters can be constrained and we are still not able to solve the component separation problem fully without any further assumptions.

We note that the form of the matrix $\mathbf{B}_\mathbf{3}$ is easy to understand from our discussions in section~\ref{subsection:Maximum Likelihood approach}: as any choice of the foreground templates is equivalent to any other, we can take any $n_c-1$ channels and define the non-CMB content of those as one of the $n_c-1$ foreground templates. These are the channels corresponding to the unit block in matrix $\mathbf{B}_\mathbf{3}$. The foreground contribution in all the other channels is a linear combination of the selected channels and can be therefore deprojected from the data. We then use the CMB prior to correct on average for the effects of the projection on the CMB signal using its assumed statistical properties. As an outcome of this procedure, in general we can estimate at best a foreground-cleaned CMB map and $n_c-1$ \textit{a priori} unknown (as they are defined by the matrix $\mathbf{M}_\mathbf{3}$) linear combinations of sky components.

As we deal only with the mixing matrix decomposition defined in Eq.~\eqref{eq:mixing matrix form}, from now on we will drop the subscript $\mathbf{3}$ to the mixing matrix for simplicity.

\subsubsection{Known CMB covariance}
\label{subsubsection:Known CMB covariance}

In this section, we assume that the CMB covariance is known. In this case, we can marginalize the likelihood in Eq.~\eqref{eq:likelihood with prior} over the CMB signal and treat it effectively as a noise source correlated between different frequency channels. The resulting total noise covariance thus becomes $\bold{\tilde{N}} \equiv \bold{A_{c}S_{c}A_{c}}^T + \bold{N}$, and the marginalized likelihood reads,
\begin{eqnarray}
	\mathcal{S} \left( \mathbf{B}_\mathbf{f}, \mathbf{s_f} | \bold{S}_\mathbf{c}\right) & = & \left( \bold{d} - \bold{B}_\mathbf{f} \bold{s_f} \right)^{T} \bold{\tilde{N}}^{-1} \left( \bold{d} - \bold{B}_\mathbf{f} \bold{s_f} \right),
\label{eq:likelihood with prior marginalized}
\end{eqnarray}
where $\mathbf{B}_\mathbf{f}$ stands for the foreground part of the mixing matrix $\mathbf{B}$, $\mathbf{s_f}$ denotes the foreground components and the likelihood is defined up to a constant that includes all the terms which do not depend on the mixing matrix and components. From this likelihood we can derive a spectral likelihood as in Eq.~\eqref{eq:genericSpectralLikelihood} which now reads,
\begin{equation}
\mathcal{S}_{\mathrm{spec}} \left( \mathbf{B}_\mathbf{f} | \bold{S}_\mathbf{c} \right) = - \bold{d}^{T} \bold{\tilde{N}}^{-1} \bold{B}_\mathbf{f} \left( \bold{B}_\mathbf{f}^{T}\bold{\tilde{N}}^{-1}\bold{B}_\mathbf{f} \right)^{-1} \bold{B}_\mathbf{f}^{T}\bold{\tilde{N}}^{-1} \bold{d}. \label{eq:spectral likelihood with prior marginalized}
\end{equation}
This likelihood simply treats the CMB signal as noise, though correlated between the frequency maps, and the only recovered sky components are foregrounds. As long as the assumed CMB covariance is correct this spectral likelihood provides unbiased estimates of the mixing elements of the matrix $\mathbf{B}_\mathbf{f}$.  This is so, because $\mathbf{B}_\mathbf{f}$ is of the form of the matrix $\mathbf{B}_\mathbf{1}$ in Eq.~\eqref{eqn:genAbreakdown} up to rearrangement of the frequency channels. This matrix is then sufficient to estimate the cleaned CMB map.

We note that this case may be relevant for some practical applications, providing some reasonable assumptions about the covariance. For instance, neglecting the primordial $B$-mode contribution to the total $B$-mode signal or assuming that the polarization signal is dominated by the $E$ modes. In both these cases, we could expect a bias due to the mismatch between the actual and assumed covariance. Whether such biases could be tolerated may need to be validated for each separate application individually. We discuss an alternative approach in the next section.

\subsubsection{Unknown CMB covariance}
\label{subsubsection:Unknown CMB covariance}

Whenever the CMB covariance is not known and has to be estimated together with the mixing matrix elements and sky components, we need to retain all the terms in Eq.~\eqref{eq:likelihood with prior} which depend on the unknown CMB covariance. Hereafter, we assume that the CMB covariance is known down to a handful of cosmological parameters. If we proceed as before, i.e. we first perform the marginalization over the CMB amplitudes and then maximize with respect to the foreground ones, we can derive the spectral likelihood which reads, 
\begin{eqnarray}
&& \mathcal{S}_{\mathrm{spec}} \left( \mathbf{B}_\mathbf{f}, \bold{S}_\mathbf{c} \right) = 
\ln | \mathbf{\tilde N} |\,
+ \, \bold{d}^{T}\bold{\Tilde{N}}\bold{d} \nonumber \\
&& - \bold{d}^{T} \bold{\tilde{N}}^{-1} \bold{B}_\mathbf{f} \left( \bold{B}_\mathbf{f}^{T}\bold{\tilde{N}}^{-1}\bold{B}_\mathbf{f} \right)^{-1} \bold{B}_\mathbf{f}^{T}\bold{\tilde{N}}^{-1} \bold{d}.
\end{eqnarray}
This likelihood maximized over the mixing matrix parameters remains unbiased if $\mathbf{S}_\mathbf{c}$ is correct. However it is biased with respect to the CMB covariance parameters even if the mixing matrix elements are correct.
This can be seen by directly computing the derivatives of this likelihood or rewriting it as,
\begin{align}
    \mathcal{S}_{\mathrm{spec}} \left( \mathbf{B}_\mathbf{f}, \bold{S}_\mathbf{c} \right) & = \ln \left| \mathbf{S}_\mathbf{c} + \left( \bold{A_{c}}^{T} \bold{N}^{-1} \bold{A_{c}} \right)^{-1} \right| \nonumber \\
    - & \mathbf{d}^t\,\mathbf{N}^{-1}\,\mathbf{B}\,
    (\mathbf{B}^t\,\mathbf{N}^{-1}\,\mathbf{B})^{-1})^{-1}\,\mathbf{B}^T\,\mathbf{N}^{-1}\,\mathbf{d} 
     \nonumber \\
    + & \mathbf{\bar{s}}_\mathrm{c}^t\,(\mathbf{S}_c\,+\,\mathbf{E}\,(\mathbf{B}^T\,\mathbf{N}^{-1}\,\mathbf{B})^{-1}\,\mathbf{E}^T)^{-1}\,
    \mathbf{\bar{s}}_\mathrm{c}. \label{eq:specCMBmargRewritten}
\end{align}
Here, $\mathbf{B}$ incorporates both $\mathbf{A}_\mathbf{c}$ and $\mathbf{B}_\mathbf{f}$ as in Eq.~\eqref{eq:mixing matrix form}, and $\mathbf{\bar{s}}$ 
is a maximum likelihood estimate of the CMB signal, i.e.,
\begin{eqnarray}
\mathbf{\bar{s}_{c}} & \equiv & \mathbf{E}^T\,\mathbf{\bar{s}} \; = \; \mathbf{E}^T\, (\mathbf{B}^T\,\mathbf{N}^{-1}\,\mathbf{B})^{-1}\,\mathbf{B}^T\,\mathbf{N}^{-1}\,\mathbf{d}, \label{eq:CMB estimate} \ \ \ \ 
\end{eqnarray}
where $\mathbf{\bar{s}}$ is given by Eq.~\eqref{eq:skyEstim} and $\mathbf{E}$ is a linear operator such that its transpose selects the CMB amplitudes from the full vector of all components and Stokes parameters.

\begin{figure*}[!htb]
\begin{center}
\includegraphics[width=1.8\columnwidth]{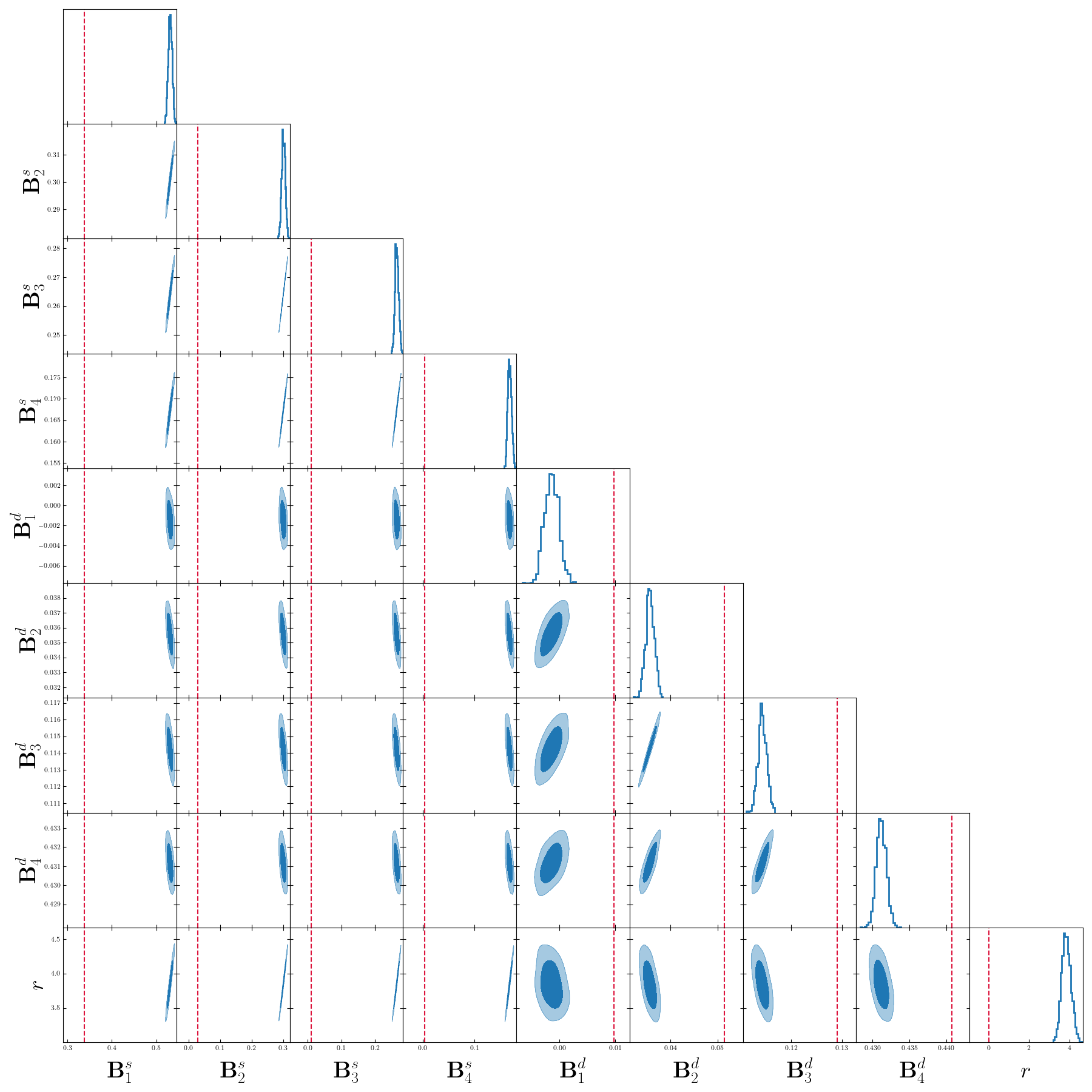}
\caption{2D and 1D densities for the likelihood in Eq.~\eqref{eq:specLikeDouble Marginalized}, i.e. in the case of unknown CMB covariance where the free parameters are} the free elements of the mixing matrix $\bold{B}$ and the tensor-to-scalar ratio $r$. The 2D densities are plotted using the \texttt{GetDist} package \cite{Lewis:2019xzd}. The mixing matrix elements are displayed in the format $\mathbf{B}_{\nu}^{c}$, where $\nu \in 0 \hdots 5$ are the frequency channels and $c \in \left( s,d \right)$ correspond to the foreground component ($s$ for synchrotron and $d$ for dust). The red points and lines correspond to the true value of the parameters used to produce the input signal, from the \texttt{PySM} foreground model \texttt{d0s0}.
\label{fig:semi-blind first try}
\end{center}
\end{figure*}

We see that this likelihood, when maximized, will in general lead to a biased estimate of the CMB covariance parameters due to the incorrect logarithmic normalization term (note that the normalization term is correct whenever $\bold{A_{c}}^{T} \bold{N}^{-1} \bold{B_{f}}=0$). This could be avoided if we reverse the order of operations starting from Eq.~\eqref{eq:likelihood with prior}, i.e. we first maximize with respect to the foreground amplitudes before marginalizing over the CMB signal. This would not affect the last two terms in Eq.~\eqref{eq:specCMBmargRewritten} but the logarithmic terms would change and the likelihood reads, 
\begin{align}
\mathcal{S}_{spec}^{\rm reverse} \left( \mathbf{B}, \bold{S}_\mathbf{c}\right)  & =\; - \mathbf{d}^T\,\mathbf{N}^{-1}\,\mathbf{B}(\mathbf{B}^T\,\mathbf{N}^{-1}\,\mathbf{B})^{-1}\,\mathbf{B}^T\,\mathbf{N}^{-1}\,\mathbf{d}\nonumber \\
+ & \mathbf{\bar{s}}^T_{\mathbf{c}}\,(\mathbf{S}_\mathbf{c}\,+\,\mathbf{E}^{T}\,(\mathbf{B}^T\,\mathbf{N}^{-1}\,\mathbf{B})^{-1}\, \mathbf{E})^{-1}\,\mathbf{\bar{s}}_{\mathbf{c}}\,\nonumber \\
+& \ln |\mathbf{S}_\mathbf{c}\,+\,\mathbf{E}^T\,(\mathbf{B}^T\,\mathbf{N}^{-1}\,\mathbf{B})^{-1}\, \mathbf{E}|\nonumber \\
- & \ln |\mathbf{E}^T\,(\mathbf{B}^T \,\mathbf{N}^{-1}\,\mathbf{B})^{-1}\,\mathbf{E}|.
\label{eq:specLikeDouble Marginalized}
\end{align}
This likelihood is manifestly unbiased for cosmological parameters when conditioned on the true values of the mixing matrix elements. However, due to explicit presence of the matrix $\mathbf{B}$ in the logarithmic terms, the likelihood for the mixing matrix elements conditioned on the true values of the CMB covariance is now biased. This highlights the fact that there is no fully consistent way of rephrasing the likelihood to satisfy both these requirements. Moreover, as we illustrate in FIG.~\ref{fig:semi-blind first try} and discuss later on, in the context of forthcoming CMB data the biases incurred in such a method are typically important and can not be ignored. 

We note that the major impact of the dependence of the logarithmic terms on the mixing matrix elements in Eq.~\eqref{eq:specLikeDouble Marginalized} is on the position of the peak rather than on its width. This suggests that if we could correct for the overall shift of the peak, approximately but in a way independent on the CMB covariance, we could get an approximately unbiased estimator with nearly correct estimates of the statistical uncertainties.

From Eq.~\eqref{eq:specLikeDouble Marginalized} we can see that when $\mathbf{S_{c}}$ is small compared to $\mathbf{N_{c}} \equiv \mathbf{E}^T \left( \mathbf{B}^T \mathbf{N}^{-1} \mathbf{B} \right)^{-1} \mathbf{E}$, both logarithmic terms would almost cancel, alleviating the problem of the bias on the mixing matrix elements. While this is not usually true in general, in some cases of interest we can recast the problem in such a way that this condition is better satisfied and at least part of the bias alleviated. Indeed, let us assume that we can split the total CMB covariance into two parts, 
\begin{eqnarray}
\mathbf{S}_\mathbf{c} \, = \, \mathbf{S}_{\mathbf{c}, \mathrm{approx}} \, + \, \Delta\mathbf{S}_\mathbf{c},
\label{eq:CMBcovSplit}
\end{eqnarray}
in a way that the first term is dominant but is independent of cosmological parameters of interest, and all the dependence on these is confined to the second, subdominant term. In this case the first term can be incorporated
as an additional contribution to the noise term,
following the procedure described in Section~\ref{subsubsection:Known CMB covariance}. The second term in Eq.~(\ref{eq:CMBcovSplit}) describing the remaining unknown CMB covariance stands for the signal covariance in Eq.~(\ref{eq:specLikeDouble Marginalized}). As this second term is, by assumption, subdominant with respect to the first one, it is also subdominant compared to the redefined noise and we can expect the biases incurred in such a procedure to be greatly suppressed as compared to the original version. While this procedure can be applied generally, it is particularly well-motivated in the case of the CMB $B$-mode data. In this case, the known part of the covariance $\mathbf{S}_{\mathbf{c}, \mathrm{approx}}$ can be straightforwardly identified with the contribution from
the lensing $B$-mode signal, and the extra variance would correspond
to the unknown primordial $B$ modes. This is the approach we follow in Sections~\ref{section:Implementation of the spectral likelihood} and \ref{section:Performances of the method}.
More general applications of this procedure will be described in subsequent work.

Denoting the effective noise as $\mathbf{\bar{N}} \, \equiv \, \mathbf{A}_\mathbf{c}\,\mathbf{S}_{\mathbf{c}, \mathrm{approx}}\,\mathbf{A}_\mathbf{c}^T \, +\, \mathbf{N}$, the proposed likelihood then reads,
\begin{align}
\mathcal{S}_{spec}^{\rm modified} \left( \mathbf{B}, \bold{S}_\mathbf{c}\right)  & = \;  - \mathbf{d}^T\,\mathbf{\bar{N}}^{-1}\,\mathbf{B}(\mathbf{B}^T\,\mathbf{N}^{-1}\,\mathbf{B})^{-1}\,\mathbf{B}^T\,\mathbf{\bar{N}}^{-1}\,\mathbf{d}\nonumber \\
+ & \mathbf{\bar{s}}^T_{\mathbf{c}}\,(\Delta\mathbf{S}_\mathbf{c}\,+\,\mathbf{E}^t\,(\mathbf{B}^T\,\mathbf{\bar{N}}^{-1}\,\mathbf{B})^{-1}\, \mathbf{E})^{-1}\,\mathbf{\bar{s}}_{\mathbf{c}}\,\nonumber \\
 +& \ln |\Delta \mathbf{S}_\mathbf{c}\,+\,\mathbf{E}^T\,(\mathbf{B}^T\,\mathbf{\bar{N}}^{-1}\,\mathbf{B})^{-1}\, \mathbf{E}| \nonumber\\
 - & \ln | \mathbf{E}^T\,(\mathbf{B}^{T}\,\mathbf{\bar{N}}^{-1}\,\mathbf{B})^{-1}\,\mathbf{E}|. \label{eq:specLikeCleaningBiased}
\end{align}

The proposed approach now involves three steps which need to be performed to derive the final equation from the original Gaussian likelihood. We first marginalize over the part of the CMB with the known covariance, then maximize with respect to the foreground component signals, to finally marginalize over the remaining part of the CMB with unknown covariance.

We note that all these difficulties could be avoided entirely if some assumptions about statistical properties of all sky components were available. For instance, it can be shown that if the foreground components were Gaussian with known covariances, one could straightforwardly derive an unbiased likelihood with respect to all parameters by simply marginalizing over all component amplitudes. This indeed has been studied in~\cite{Cardoso:2008qt,Vansyngel:2014dfa} for isotropic and stationary foregrounds. In such cases, we could solve the component separation problem completely. However, such assumptions are in general incorrect and their potentially important impact on the final results is often difficult to quantify. We note, were the foregrounds truly Gaussian, we still would need to know their correct covariances to avoid biases. The aim of the proposed approach is more modest, as we attempt to clean only the CMB map of the foregrounds, while avoiding such strong assumptions and retaining some level of understanding and control of the resulting biases.

\subsubsection{Relation to the template fitting}

If we can assume that $n_c-1$ of the frequency channels are each dominated by a different sky component,
then the mixing matrix would be given by a matrix of the form of Eq.~\eqref{eq:mixing matrix form}, up to a reordering of its rows. Consequently, the transformation matrix $\mathbf{M}$ would be forced to be equal to the identity. We could then use the proposed procedure to perform an actual component separation where the output estimates of the sky would correspond to actual physical sky components. This procedure is usually referred to as template fitting \cite{Efstathiou:2009MNRAS.397.1355E, Katayama:2011ApJ...737...78K,Ichiki:2018glx}. In a typical application, it is assumed that two extreme frequency channels are dominated by the synchrotron (lowest frequency) and thermal dust emission (highest frequency) emissions. The procedure proposed here, because it assumes the same form for the mixing matrix, can be looked at as a generalization of such a technique applicable to the cases when such assumptions are not warranted. This would be the case, for instance, when the number of relevant components exceeds two and not all of them dominate one of the frequency bands, or when the covered frequency range is too narrow, etc.

\subsubsection{Relation to Internal Linear Combination}

Let us start from Eq.~\eqref{eq:specLikeDouble Marginalized} but assume that the CMB covariance is known. In this case, the logarithmic terms of the likelihood are not needed because we are not fitting for the parameters of $\mathbf{S_{c}}$, and we can decide to drop them.
The remaining terms of the spectral likelihood can then be recast as,
\begin{align}
\mathcal{S}_{spec} \left( \mathbf{B} | \bold{S}_\mathbf{c}\right)  & = - \mathbf{d}^T \mathbf{N}^{-1} \mathbf{B} \left( \mathbf{B}^{T} \mathbf{N}^{-1} \mathbf{B} \right)^{-1} \mathbf{B}^T \mathbf{N}^{-1} \mathbf{d} \nonumber \\
+ \mathbf{\Bar{s}}^{T}_\mathbf{c} & \left( \mathbf{S}_\mathbf{c} + \mathbf{E}^{T} \left( \mathbf{B}^T \mathbf{N}^{-1} \mathbf{B} \right)^{-1} \mathbf{E} \right)^{-1}\mathbf{\Bar{s}}_\mathbf{c},
\label{eq:specLikeILCstep1}
\end{align}
Because the ILC method is expected to clean the foreground by a weighted linear combination of frequency observations, it explicitly or implicitly assumes that there are as many components as weights, i.e. as many components as the number of frequency channels. In this case, the mixing matrix is a square matrix and the first term in the likelihood reduces to the constant term $\mathbf{d}^T \mathbf{N}^{-1} \mathbf{d}$, leaving only the second term varying. Given that the mixing matrix $\mathbf{B}^{s}_\mathbf{p}$ is now a square and invertible matrix, its unknown elements are only those contained in its first row, and they can be recast in a form of a weight matrix $\mathbf{W}$
defined as,
\begin{eqnarray}
\mathbf{W}^T & \equiv &  \mathbf{E}^{T}\,(\mathbf{B}^T\,\mathbf{N}^{-1}\,\mathbf{B})^{-1}\,\mathbf{B}^T\,\mathbf{N}^{-1}\; = \;  \mathbf{E}^{T} \mathbf{B}^{-1}.
\label{eq:ILCweights}
\end{eqnarray}

We note that the sum of elements of $\mathbf{W}$ for a given Stokes parameter and a given pixel has to be equal to 1, i.e.,
\begin{eqnarray}
    \left( \ps{W} \right)^{T} \bold{e} = 1,
     \label{eq:weightConstILC}
\end{eqnarray}
where $\mathbf{e}$ is a vector of ones of length $n_c$. This reflects the fact that the first, i.e. CMB, column of the mixing matrix $\mathbf{B}$ is assumed to be set to $\mathbf{A}_\mathbf{c} = \mathbf{e}$, meaning that single frequency maps are expressed in CMB units and that the CMB frequency dependence is known \textit{a priori}.

With all these simplifications, the likelihood can be rewritten as,
\begin{align}
\mathcal{S}_{spec} \left( \mathbf{W} | \bold{S}_\mathbf{c}\right)  & =  \; \mathbf{\Bar{s}}^T_{\mathbf{c}}\,(\mathbf{S}_\mathbf{c}\,+\,\mathbf{W}\,\mathbf{N}\, \mathbf{W}^T)^{-1}\,\mathbf{\Bar{s}}_{\mathbf{c}}\nonumber\\
=  \; & \mathbf{d}^T\,\mathbf{W}^T\,(\mathbf{S}_\mathbf{c}\,+\,\mathbf{W}\,\mathbf{N}\, \mathbf{W}^T)^{-1}\,\mathbf{W}\,\mathbf{d}.
\label{eq:specLikeILCstep3}
\end{align}
We can write this likelihood directly in the harmonic domain, introducing different weights for different value of $\ell$, and assuming that the noise is isotropic. For simplicity, we restrict ourselves to the cases when the CMB covariance is diagonal, i.e., either the case of total intensity or the polarization-only (with no parity violation). The corresponding likelihood then reads,
\begin{align}
\mathcal{S}_{spec} \left( \mathbf{w}_\ell | \mathcal{C}_\ell\right)  & 
=  \; \sum_{\ell, m}\, \mathbf{w}_\ell\,(\mathcal{C}_\ell\,+\,\mathbf{w}_\ell\,\sigma_\ell^2\, \mathbf{w}_\ell)^{-1}\,\mathbf{w}_\ell\;\mathbf{d}_{\ell m}\mathbf{d}^\dagger_{\ell m}.
\label{eq:specLikeILCstep4}
\end{align}
We can now use the above likelihood to maximize with respect to $\mathbf{w}_\ell$, subject to the constraint as in Eq.~\eqref{eq:weightConstILC} using Lagrange multipliers. For all the modes for which the noise is subdominant we obtain,
\begin{equation}
    \mathbf{w}_\ell = \frac{\displaystyle \big( \sum_m\;\mathbf{d}_{\ell m}\mathbf{d}^\dagger_{\ell m}\big)^{-1}\, \mathbf{e}}{\displaystyle \mathbf{e}^T\,\big( \sum_m\;\mathbf{d}_{\ell m}\mathbf{d}^\dagger_{\ell m}\big)^{-1}\, \mathbf{e}},
\end{equation}
\begin{equation}
    \mathbf{s}_{\mathbf{c}, \ell, m} = \mathbf{w}_\ell^T\,\mathbf{d}_{\ell m} \; = \; \frac{\displaystyle \mathbf{e}^T\,\big( \sum_m\;\mathbf{d}_{\ell m}\mathbf{d}^\dagger_{\ell m}\big)^{-1}}{\displaystyle \mathbf{e}^T\,\big( \sum_m\;\mathbf{d}_{\ell m}\mathbf{d}^\dagger_{\ell m}\big)^{-1}\, \mathbf{e}}\mathbf{d}_{\ell m},
\end{equation}
which are essentially the ILC estimates. If the same weights are assumed for all harmonic modes or some subset of those denoted $\mathbf{L}$ where the noise is subdominant, then the maximization leads instead to,
\begin{eqnarray}
\mathbf{w}  & = & \frac{\displaystyle \bigg( \sum_{\ell \in \mathbf{L}, m}\;\mathcal{C}_\ell^{-1}\,\mathbf{d}_{\ell m}\mathbf{d}^\dagger_{\ell m}\bigg)^{-1}\, \mathbf{e}}{\displaystyle \mathbf{e}^T\,\bigg( \sum_{\ell \in \mathbf{L}, m}\;\mathcal{C}_\ell^{-1}\,\mathbf{d}_{\ell m}\mathbf{d}^\dagger_{\ell m}\bigg)^{-1}\, \mathbf{e}},
\end{eqnarray}
and the contribution from different harmonic modes of the data is weighted according to the assumed CMB power spectrum. As this is in principle unknown one may opt for using some best fit approximation instead somewhat similar to our considerations in Sect.~\ref{subsubsection:Unknown CMB covariance}. However, one can expect the weights to be nearly unbiased only if the approximation is good enough. 

In contrast, the method proposed in Sect.~\ref{subsubsection:Unknown CMB covariance} allows for a proper weighting of all the modes, even in the cases when
CMB is not dominant and for a consistent estimation of the \textit{a priori} unknown CMB power spectrum.

\begin{figure*}[!htb]
\begin{center}
\includegraphics[width=1.8\columnwidth]{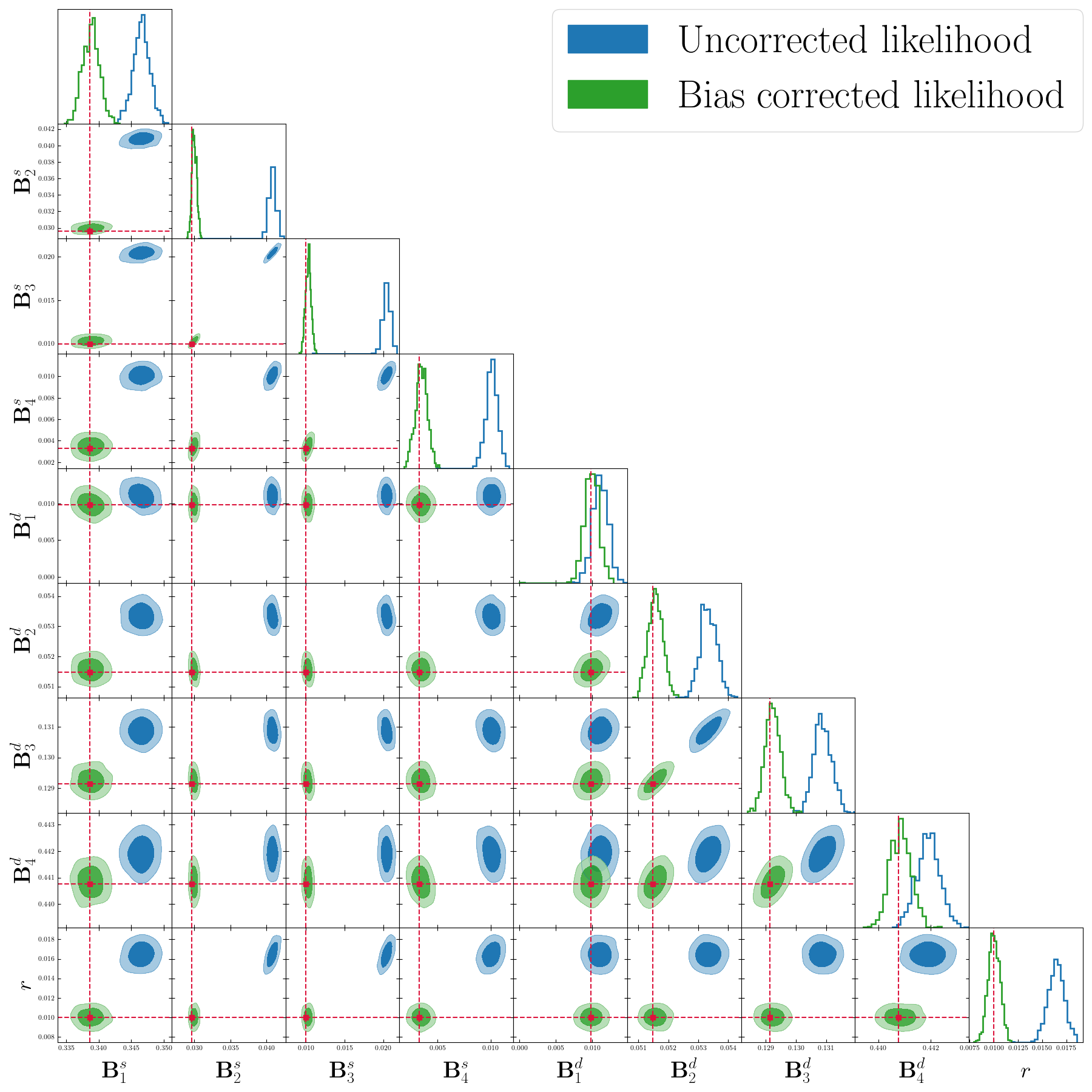}
\caption{2D and 1D densities for the free parameters of the likelihoods in Eq.~\eqref{eq:specLikeDouble Marginalized} (blue) and Eq.~\eqref{eq:likelihood lensing as noise} (green), i.e. the free elements of the mixing matrix $\bold{B}$ and the tensor-to-scalar ratio $r$. The 2D densities are plotted using the \texttt{GetDist} package \cite{Lewis:2019xzd}. The mixing matrix elements are displayed in the format $\mathbf{B}_{\nu}^{c}$, where $\nu \in 0 \hdots 5$ are the frequency channels and $c \in s,d$ correspond to the foreground component ($s$ for synchrotron and $d$ for dust). The red points and lines correspond to the true value of the parameters used as input for the input signal. In order to mitigate the bias (see discussion in Section~\ref{section:Implementation of the spectral likelihood}), the evaluation of the likelihood in Eq.~\eqref{eq:specLikeDouble Marginalized} is restricted to multipoles lower than $\ell_{\rm max} = 150$. The evaluation of the likelihood in Eq.~\eqref{eq:likelihood lensing as noise} goes up to the full band limit of $\ell_{\rm max} = 256$. In all cases, the model of foregrounds used is \texttt{d0s0}.}
\label{fig:comparison CMB prior and CMB primordial prior}
\end{center}
\end{figure*}

\section{Implementation of the spectral likelihood}
\label{section:Implementation of the spectral likelihood}

In this section we demonstrate the proposed method and validate our analytic expectations from the previous sections, in particular those concerning the expected biases on the parameters. To allow for an efficient exploration of different features of the method we will rely on some simplifying assumptions and idealizations, and leave its full implementation to future work.

Consequently, we work hereafter in the harmonic domain as the covariance of CMB signal is most conveniently expressed in the spherical harmonics basis. Moreover, we focus on large scale and use multipoles up to $\ell_{\rm max} = 256$ unless stated otherwise, as these are the most relevant for measuring the tensor-to-scalar ratio, $r$. For this same reason, we also assume that only the $B$-mode polarization data are available and discard any information from total intensity and $E$ modes of polarization. Therefore, the CMB covariance for the multipole $\ell$ is, in our case, just the CMB $B$-mode angular power spectrum, that we can express as:
\begin{eqnarray}
    \left( \bold{S_{c}} \right)_{\ell} & = & C_{\ell}^{BB} =  r C_{\ell}^{BB, \rm prim} + C_{\ell}^{BB, \rm lens} \\
    & \equiv & \left( \bold{S_{prim}} \right)_{\ell} + \left( \bold{S_{lens}} \right)_{\ell}.
\end{eqnarray}
Here $C_{\ell}^{BB, \rm prim}$ is the primordial contribution to $B$ modes, $C_{\ell}^{BB, \rm lens}$ is the contribution from gravitationally lensed E modes, and $r$ is the tensor-to-scalar ratio which is assumed to be the only free parameter of the CMB covariance for this study. As the measurement of $r$ is an important goal of current and future CMB experiments, we take its accurate reconstruction as our figure of merit in the following. Similarly, the noise covariance is given by its angular power spectrum $\mathbf{N}_{\ell}$, and we assume a frequency-channel dependent white noise in the following. The level of this white noise is determined by the experimental set-up under consideration. To start with we choose an experimental set-up with relatively few frequency channels to limit the number of unknown parameters. The specific set-up we consider is based on SO-SAT characteristics \cite{SimonsObservatory:2018koc},  which we detail in TABLE~\ref{tab:SO-SAT-like}. 

%
%

\begin{table}[!htb]
\renewcommand{\arraystretch}{2}
\begin{center}
\begin{tabular}{|P{2cm}|c|c|c|c|c|c|}
\hline
\multicolumn{7}{|c|}{SO-SAT-like} \\ \hline\hline
\multirow{1}{2cm}{\centering $\nu$\\ (GHz)} & 27 & 39 & 93 & 145 & 225 & 280 \\ \hline
\multirow{1}{2cm}{\centering Sensitivity\\ ($\mu$K-arcmin)} & 49.5 & 29.7 & 3.7 & 4.7 & 8.9 & 22.6 \\ \hline
$f_{\rm sky}$ & \multicolumn{6}{c|}{$\sim 10\%$} \\ \hline

\end{tabular}
\caption{Principal characteristics of the SO-SAT-like set-up used in this study: center frequency of the frequency bands, sensitivity in polarization and observed fraction of the sky.}
\label{tab:SO-SAT-like}
\end{center}
\end{table}

As mentioned in Section~\ref{subsubsection:Unknown CMB covariance}, we include the lensing part of the CMB signal in the noise in order to mitigate the bias. The likelihood reads in the harmonic domain,

\begin{align}
\mathcal{S}_{spec} \left( \mathbf{B}, \bold{S_{\rm prim}} \right)  & =\; - \mathbf{d}^T\,\mathbf{\tilde{N}}^{-1}\,\mathbf{B}(\mathbf{B}^T\,\mathbf{\tilde{N}}^{-1}\,\mathbf{B})^{-1}\,\mathbf{B}^T\,\mathbf{\tilde{N}}^{-1}\,\mathbf{d}\nonumber \\
+ & \mathbf{\bar{s}}^T_{\mathbf{c}}\,(\mathbf{S_{prim}}\,+\,\mathbf{E}^t\,(\mathbf{B}^T\,\mathbf{\tilde{N}}^{-1}\,\mathbf{B})^{-1}\, \mathbf{E})^{-1}\,\mathbf{\bar{s}}_{\mathbf{c}}\,\nonumber \\
+& \ln |\mathbf{S_{prim}}\,+\,\mathbf{E}^T\,(\mathbf{B}^T\,\mathbf{\tilde{N}}^{-1}\,\mathbf{B})^{-1}\, \mathbf{E}|\nonumber \\
- & \ln |\mathbf{E}^T\,(\mathbf{B}^T \,\mathbf{\tilde{N}}^{-1}\,\mathbf{B})^{-1}\,\mathbf{E}|.
\label{eq:likelihood lensing as noise}
\end{align}
and the noise covariance matrix includes the lensing power,
\begin{equation}
    \mathbf{\tilde{N}}_{\ell} = C_{\ell}^{BB, \rm lens}\bold{A_{c}A_{c}}^{T} + \bold{N}_{\ell}.
\end{equation}

Consistently with the current knowledge of the polarized foregrounds in the microwave band, see e.g. \cite{Planck:2018yye}, we assume that there are two \textit{a priori} unknown foreground components in the component separation procedure, i.e. that the mixing matrix has three columns. The sky simulations include CMB, noise and foreground templates with their SED taken from the \texttt{d0} model for Galactic dust and \texttt{s0} model for synchrotron of the \texttt{PySM} software \cite{Thorne:2016ifb,Zonca:2021row}, unless stated otherwise.

In order to explore the likelihood, we produce independent realizations of noise $a_{\ell m}$ following the white noise power spectrum, and of CMB $a_{\ell m}$ following the total CMB $B$-mode power spectrum (including primordial gravitational waves and lensing), for given values of $r$. Then we maximize the likelihood for each of these realizations. FIG.~\ref{fig:comparison CMB prior and CMB primordial prior} shows the density of these maxima in all of the free parameter 2D planes, and the corresponding 1D densities for each of these parameters, using the likelihoods in Eqs.~\eqref{eq:specLikeDouble Marginalized} and~\eqref{eq:likelihood lensing as noise}. The free parameters are the elements of the mixing matrix with the structure defined in Eq.~\eqref{eq:mixing matrix form}. We are free to choose, by rearranging the lines of the mixing matrix, two frequency channels that have their elements fixed. From the known characteristics of the two foreground components, we choose to fix the elements of the lowest and highest frequency channels. This corresponds to identifying one of the components with synchrotron emission, plus residual subdominant dust, and the other with Galactic dust, plus residual subdominant synchrotron. In addition to making physical sense, we have found that the Fisher information matrix of the likelihood is in general better conditioned when choosing to fix the two extreme frequency channels.

We see that the reconstructed parameters are severely biased for the likelihood in Eq.~\eqref{eq:specLikeDouble Marginalized}, but not for the likelihood in Eq.~\eqref{eq:likelihood lensing as noise}, as expected from the discussion in Section~\ref{subsubsection:Unknown CMB covariance}.

One major advantage of the maximum likelihood approach over other existing approaches such as ILC is that we can estimate \textit{a priori}, at least qualitatively, the level of bias. In particular, we can estimate the contribution of each multipole to the overall bias if we assume that the noise and CMB average of the spectral likelihood  as a function of the elements of the mixing matrix $\beta$ is Gaussian:
\begin{equation}
    \left< \mathcal{S}_{\rm spec} \left( \beta | r \right) \right> \simeq \mathrm{const} + \left( \beta - \bar{\beta} \right)^{T} \mathcal{B}^{-1} \left( \beta - \bar{\beta} \right),
\end{equation}
where $\bar{\beta}$ is the location of the likelihood's peak and $\mathcal{B}$ is the covariance matrix of the gaussian likelihood for the $\beta$ parameters. We want to find the bias on the spectral parameters, i.e. $\delta \beta = \beta_{\rm true} - \bar{\beta}$, which can be expressed as:
\begin{equation}
\delta \beta = \left< \frac{\partial^{2} \mathcal{S}}{\partial \beta \partial \beta'} \right>^{-1} \left. \left< \frac{\partial \mathcal{S}}{\partial \beta} \right> \right|_{\beta_{\rm true}} \equiv \frac{1}{2} \mathcal{F}^{-1} \left. \left< \frac{\partial \mathcal{S}}{\partial \beta} \right> \right|_{\beta_{\rm true}},
\end{equation}
where $\mathcal{F}$ is the Fisher matrix of the likelihood. Therefore, the contribution from the multipole $\ell$ to the bias is defined to be:
\begin{equation}
    \delta \beta_{\ell} = \frac{1}{2} \mathcal{F}^{-1} \left. \left< \frac{\partial \mathcal{S}_{\ell}}{\partial \beta} \right> \right|_{\beta_{\rm true}} \quad \text{with} \quad \mathcal{S}_{\rm spec} = \sum_{\ell} \mathcal{S}_{\ell}. \label{eq:bias per ell}
\end{equation}

\begin{figure}[!htb]
\begin{center}
\includegraphics[width=\columnwidth]{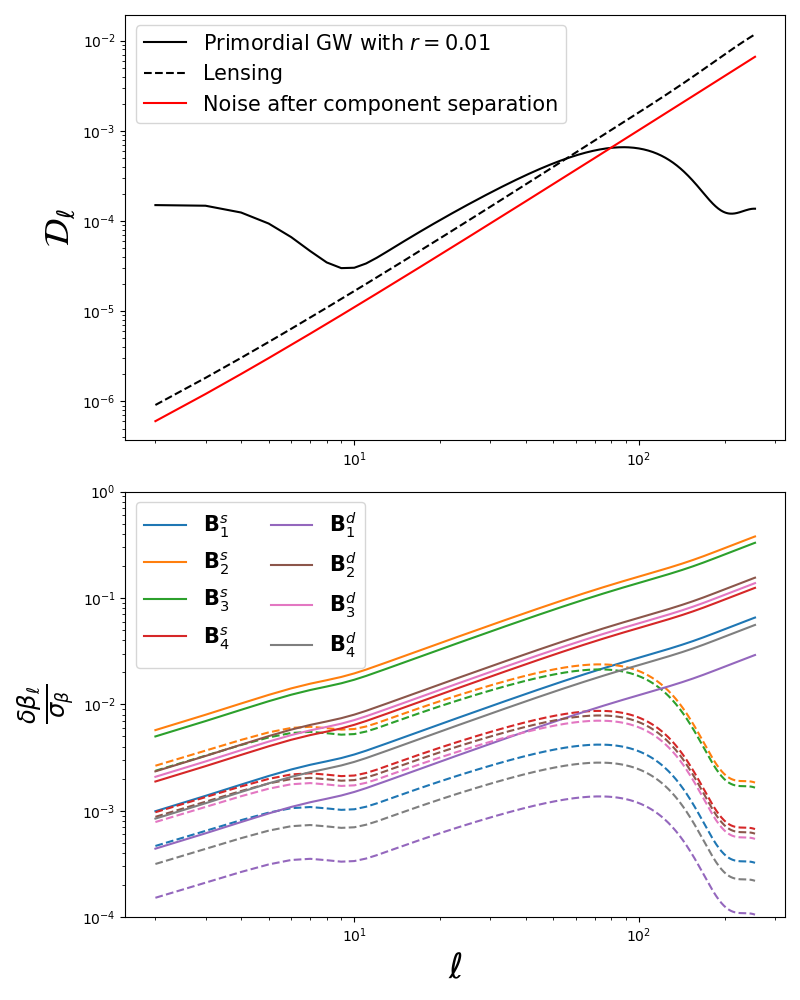}
\caption{(Top) Power-spectra of the different elements entering the computation of the bias, from Eq.~\eqref{eq:bias per ell}. (Bottom) Contribution to the bias as a function of the multipole, relatively to the statistical error $\sigma_{\beta}$ computed from the Fisher matrix, for each of the free elements of the mixing matrix for the likelihoods in Eq.~\eqref{eq:specLikeDouble Marginalized} (solid) and in Eq.~\eqref{eq:specLikeCleaningBiased} (dashed).}
\label{fig:bias vs ell}
\end{center}
\end{figure}

The contribution to the bias as a function of the multipole, relatively to the statistical error $\sigma_{\beta}$ defined as the square root of the diagonal elements of $\mathcal{F}^{-1}$, is depicted in FIG.~\ref{fig:bias vs ell} for the two likelihoods.

It shows that the parameters which are the most biased are those in the central frequency bands, at 93 GHz and 145 GHz, where the CMB signal is the most relevant. This explains why the bias on the $\beta$ parameters are able to propagate and induce a significant bias on $r$. This confirms the results shown in FIG.~\ref{fig:comparison CMB prior and CMB primordial prior}.

In order to understand this behaviour as a function of $\ell$, we need to compute the derivatives of the spectral likelihood which enter in the computation of $\delta \beta$. Generically, these derivatives can be put in the form,
\begin{equation}
    \left< \frac{\partial \mathcal{S}_{\rm spec}}{\partial \beta} \right> = \mathrm{Tr} \left[ \left( \left( \bold{S_{c}} + \bold{N}\bold{_{c}} \right)^{-1} - \bold{N}\bold{_{c}}^{-1} \right) \bold{N}\bold{_{c}}_{,\beta} \right], \label{eq:bias}
\end{equation}
where $\mathbf{S_{c}}$ is either the full CMB $B$-mode signal covariance or only that of the primordial GW signal, and $\mathbf{N_{c}}$ is the noise after component separation either not including the lensing $B$-mode signal or including it. The subscript $_{, \beta}$ indicates a derivatives with respect to the mixing matrix parameters. Several ingredients enter in the expression of these derivatives: the CMB signal covariance, elements of the mixing matrix and the noise covariance. The two former elements are derived from the method, while the latter is defined \textit{a priori}. Therefore, it is interesting to consider the asymptotic behaviour when the noise level goes to the limit by introducing an affine parameter $\lambda$ and defining $\bold{N} = \lambda \bold{N}_{0}$ while $\bold{N}_{0}$ is kept constant:
\begin{eqnarray}
    \left< \frac{\partial \mathcal{S}_{\rm spec}}{\partial \beta} \right> & \xrightarrow[\lambda \rightarrow 0]{} & - \mathrm{Tr} \left[ \left( \bold{N}\bold{_{c}}^{0} \right)^{-1} \bold{N}\bold{_{c}}^{0}_{,\beta} \right], \\
    & \xrightarrow[\lambda \rightarrow \infty]{} & 0, \label{eq:asymptotic bias}
\end{eqnarray}
with $\bold{N}\bold{_{c}}^{0} = \bold{E}^{T} \left( \bold{B}^{T} \bold{N}_{0}^{-1} \bold{B} \right)^{-1} \bold{E}$.

\begin{figure*}[!thb]
\begin{center}
\includegraphics[width=1.5\columnwidth]{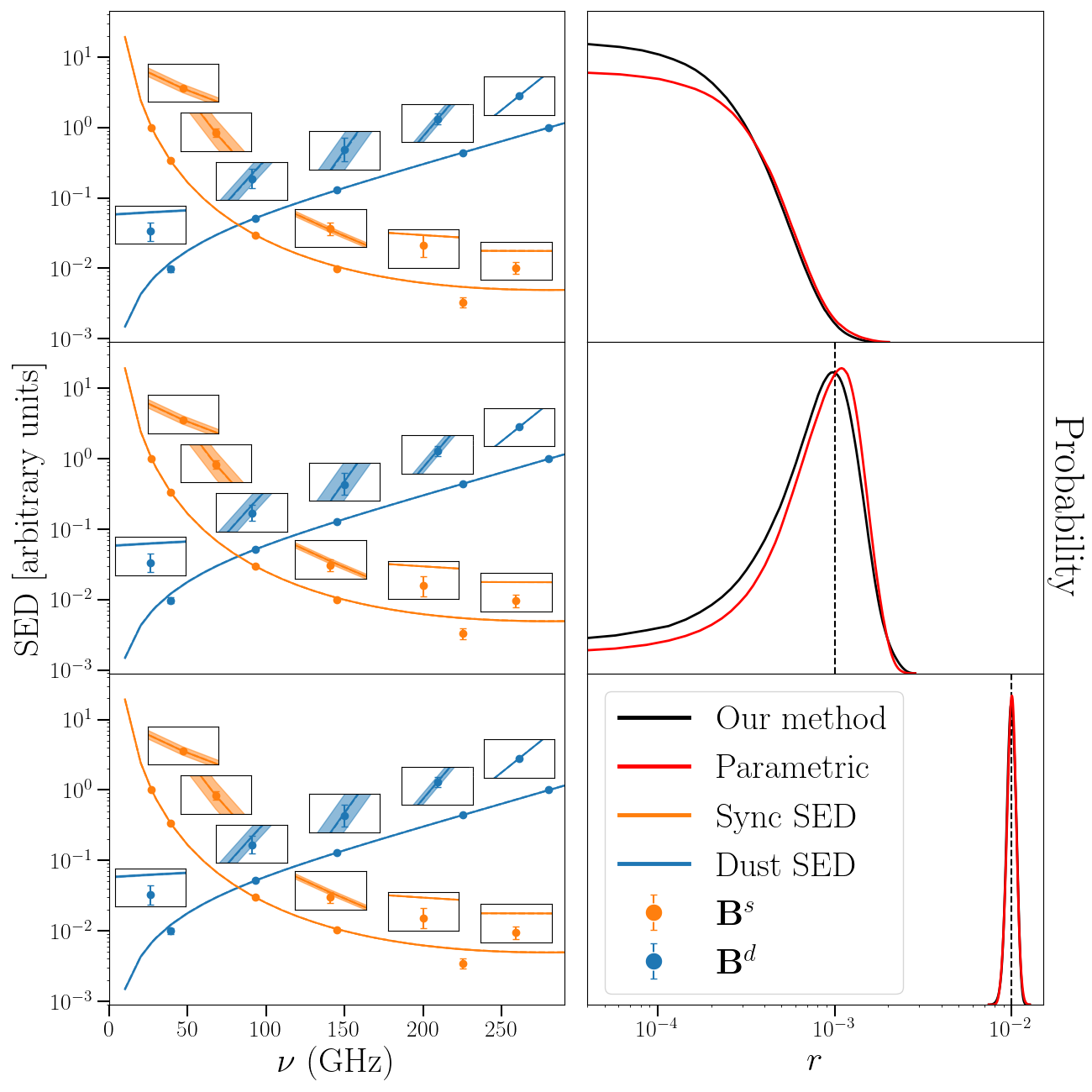}
\caption{Comparison of the constraints on the spectral parameters (Left) and tensor-to-scalar ratio (Right) in the SO-SAT-like case when the true value of $r$ is 0.01 (Top), 0.001 (Middle) and 0 (Bottom) from the standard parametric approach of \texttt{FGBuster} and the method described in the present paper. The dots with error bars correspond to the mean and 68\% confidence interval of the marginal posterior of the mixing matrix elements from our method, the mean of the SEDs from the parametric approach correspond to the dashed lines, overlayed with the true SEDs represented by the solid lines following the model of foregrounds \texttt{d0s0}, and the 68\% confidence intervals are represented by the shaded areas.}
\label{fig:results SO-SAT-like d0s0}
\end{center}
\end{figure*}

Therefore, as already mentioned, this method is unbiased in the limit of vanishing signal, in other words for low signal-to-noise ratio (SNR), as defined by $\bold{S_{c}}\bold{N}\bold{_{c}}^{-1}$. But, of course, this comes at the cost of statistical uncertainties on the parameters, that grow  with the noise level. Including the lensing in the noise regularizes the problem, ensuring that the SNR is low without affecting the statistical uncertainties as was noted in Section~\ref{subsubsection:Unknown CMB covariance}

\section{Performances of the method}
\label{section:Performances of the method}

From FIG.~\ref{fig:comparison CMB prior and CMB primordial prior}, we see that there is at least one case where the proposed method is able to properly estimate the tensor-to-scalar ratio $r$, which constitutes our figure of merit. The goal of this section is to apply this method to three test cases in order to explore its potential domain of application. For reference, we compare its performances with a standard parametric method, as implemented in the \texttt{FGBuster} package \cite{Stompor:2008sf,Stompor:2016hhw}. We restrict ourselves to a simple implementation of this parametric approach in the harmonic domain, so both approaches are applied on similar footings. In particular, none of the implementations can mitigate spatial variations of the foreground SEDs \cite{Errard:2018ctl}. More advanced implementations of the parametric method which efficiently deal with such effects are already available, and similar extensions can be used in principle in our approach. This requires significant additional work and is left here for the future. Therefore, we emphasize that the results discussed hereafter do not pertain to being representative of what can be achieved using a parametric method, but provide merely a point of reference.

\subsection{SO-SAT-like}

The first test case to which we apply our method is the implementation described in Section \ref{section:Implementation of the spectral likelihood}, of which FIG.~\ref{fig:comparison CMB prior and CMB primordial prior} shows the first result. We consider three cases with different values for the true tensor-to-scalar ratio: \\
$r = 0, 0.001 \text{ and } 0.01$.

We compare the results using our method with the parametric approach from \texttt{FGBuster}, using a Modified Black-Body (MBB) SED for the dust component, with spectral parameter $\beta_{d}$ and temperature $T_{d}$, and a Power-Law SED for synchrotron emission, with spectral parameter $\beta_{s}$, i.e. the same models as those used to produce the foreground maps with \texttt{PySM}. Following the specifications from the SO collaboration \cite{Wolz:2023lzb}, we do not vary the temperature of the MBB and set it to $T_{d} = 20 \mathrm{\ K}$.

For easier visualization of the results on the free parameters, and comparison with the results from the parametric method, we separate them into constraints on the spectral parameters and the marginal posterior for $r$. In FIG.~\ref{fig:results SO-SAT-like d0s0}, the constraints on the spectral parameters are presented in the form of SEDs where we identify the elements of the first column of $\bold{A_{f}}$ with the SED of synchrotron emissions and the elements of the second column with the SED of thermal dust, as would be the case with a template fitting approach. This allows for a comparison with the results on spectral parameters from the parametric method, which correspond to an SED with statistical uncertainties.

In addition to very good performances in recovering $r$, we see an impressive agreement between the true SED used in the simulations and the corresponding matrix elements from our method, except at extreme frequencies. A potential reason for this slight disagreement is that the reconstructed components are not exactly synchrotron and dust. Indeed, they are defined to be the complete foreground contribution at the lowest and highest frequencies, respectively, and are therefore slightly contaminated by the other foreground component. Nevertheless, the SEDs are accurately reproduced.

\begin{figure}[!htb]
\begin{center}
\includegraphics[width=\columnwidth]{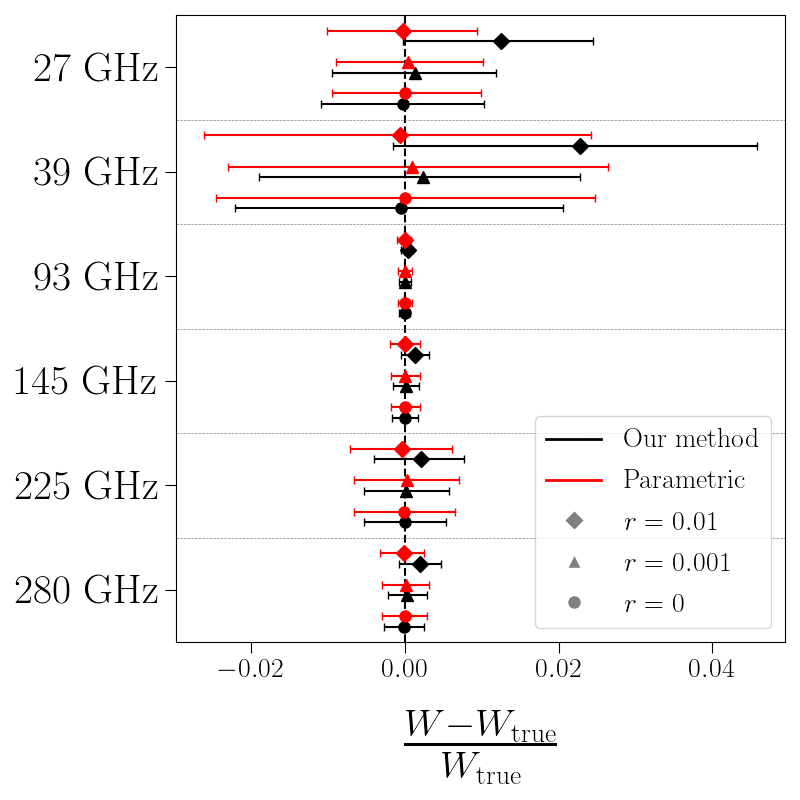}
\caption{Mean and 68\% confidence interval of the relative weights, compared with the true weights used as input on the simulations, for each of the frequency channels of the SO-SAT-like case, for the three cases $r = 0, 0.001 \text{ and } 0.01$. In all cases, the model of foregrounds used as reference is \texttt{d0s0}.}
\label{fig:weights SO d0s0}
\end{center}
\end{figure}

The performances of the component separation itself can be assessed by looking at the weights used to produce the CMB map in Eq.~\eqref{eq:CMB estimate}, $\bold{\Bar{s}_{c}} \equiv \bold{W}^{T} \bold{d}$. Here, we use the CMB estimate defined from $\bold{N}$ instead of $\bold{\tilde{N}}$. Indeed, using $\bold{\tilde{N}}$ would correspond to a Wiener filtered estimate of the CMB map, while the generalized least-square estimate \eqref{eq:CMB estimate} gives the best cleaned CMB map. The mean and 68\% confidence interval of the weights obtained for the three cases studied in this section are presented in FIG.~\ref{fig:weights SO d0s0}, as compared with the true weights obtained from the true scaling laws used to produce the simulations. Their marginal distributions are obtained by computing \textit{a posteriori} the weights for the maximum likelihood parameters of each of the CMB and noise realizations.

We see that the frequency maps are correctly weighted to reconstruct the CMB map, with a slight offset, in the 1-$\sigma$ range, observed for extreme frequency channels when $r=0.01$. This offset vanishes for lower values of $r$ as is expected because the SNR gets lower in this case and \eqref{eq:asymptotic bias} tells us that the bias on the parameters gets suppressed. Surprisingly enough, the statistical uncertainties on the weights are slightly smaller in our method than with the parametric approach which is fitting parameters of the exact same SED as the one used in the simulations.

\begin{figure*}[!htb]
\begin{center}
\includegraphics[width=1.5\columnwidth]{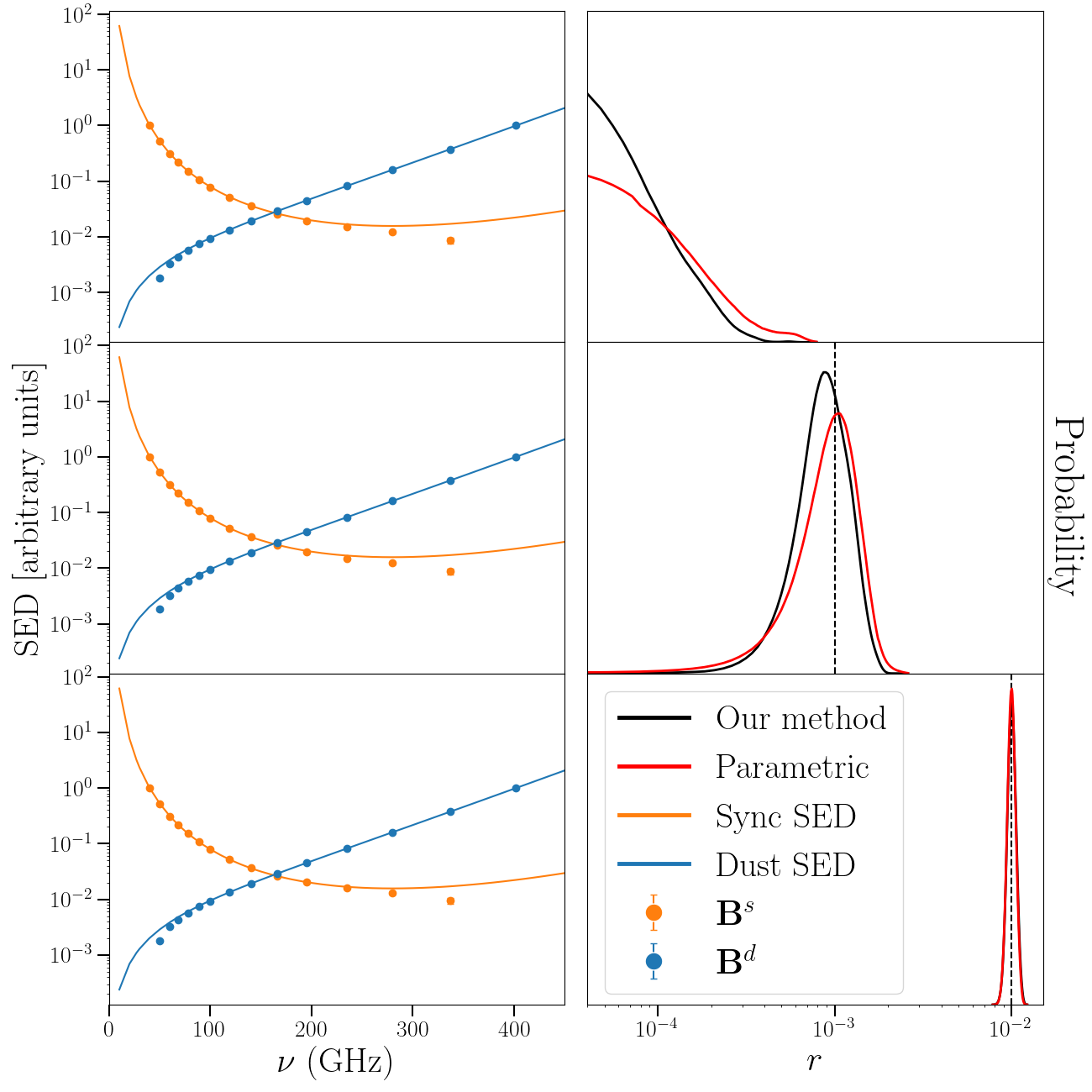}
\caption{Constraints on the mixing matrix free elements in the LiteBIRD-like case with our approach (Left) and comparison of tensor-to-scalar ratio posterior with the standard parametric approach (Right) when the true value of $r$ is 0.01 (Top), 0.001 (Middle) and 0 (Bottom). The dots with error bars correspond to the mean and 68\% confidence interval of the marginal posterior of the mixing matrix elements from our method and the solid lines represent the true SEDs following the model of foregrounds \texttt{d0s0}.}
\label{fig:results LiteBIRD-like d0s0}
\end{center}
\end{figure*}

\subsection{LiteBIRD-like}

Since the number of parameters grows with the number of frequency channels in our case, we need to test the performances of the method when the number of frequency channels is larger. Therefore, we test the framework on a LiteBIRD-like experimental case that has more than double the number of frequency channels, with characteristics described in TABLE~\ref{tab:LiteBIRD-like}~\cite{Hazumi:2019lys}.

\begin{table}[!htb]
\renewcommand{\arraystretch}{2}
\begin{adjustbox}{width=\columnwidth,center}
\begin{tabular}{|c|c||c|c|}
\hline
\multicolumn{4}{|c|}{LiteBIRD-like} \\ \hline\hline
\multirow{1}{2cm}{\centering $\nu$\\ (GHz)} & \multirow{1}{2cm}{\centering Sensitivity\\ ($\mu$K-arcmin)} & \multirow{1}{2cm}{\centering $\nu$\\ (GHz)} & \multirow{1}{2cm}{\centering Sensitivity\\ ($\mu$K-arcmin)} \\ \hline
40 & 37.5 & 140 & 5.9 \\ \hline
50 & 24.0 & 166 & 6.5 \\ \hline
60 & 19.9 & 195 & 5.8 \\ \hline
68 & 16.2 & 235 & 7.7 \\ \hline
78 & 13.5 & 280 & 13.2 \\ \hline
89 & 11.7 & 337 & 19.5 \\ \hline
100 & 9.2 & 402 & 37.5 \\ \hline
119 & 7.6 \\ \hline
$f_{\rm sky}$ & \multicolumn{3}{c|}{$\sim 50\%$} \\ \hline
\end{tabular}
\end{adjustbox}
\caption{Principal characteristics of the LiteBIRD-like set-up used in this study: center frequency of the frequency bands, sensitivity in polarization and observed fraction of the sky. LiteBIRD will observe the entirety of the sky but we use the Planck HFI mask \cite{PLA} to hide the Galactic plane, leaving an effective sky fraction of approximately half of the sky.}
\label{tab:LiteBIRD-like}
\end{table}

Because LiteBIRD has more frequency channels, it has more power in constraining the SEDs of foreground components. Therefore, in this case we release the temperature of dust $T_{d}$ in the parametric approach, following the recipes from the LiteBIRD collaboration \cite{LiteBIRD:2022cnt}.

As the number of free elements of the mixing matrix goes from 8 in the SO-SAT-like case to 26 in the LiteBIRD-like case, it would not be surprising if the good performances we witnessed in the previous section would break down in the latter case. In particular, we would expect the statistical uncertainty on $r$ to be significantly larger than in the parametric alternative, just from the larger number of parameters. However, we can see in FIG.~\ref{fig:results LiteBIRD-like d0s0} that this is not the case, and the statistical uncertainty on $r$ is actually smaller than in the parametric case for low values of $r$. The propagation of all the partial uncertainties to the final uncertainty on $r$ is not trivial but these results suggest that, in the studied cases, the impact of the uncertainties on the mixing elements is subdominant while the impact of noise is different due to different weights assigned by both methods to different frequency channels. These observations are only tentative and can be expected to be very case dependent. Their full exploration  will require a more advanced implementation of the method and will be presented in future work.

\begin{figure}[!htb]
\begin{center}
\includegraphics[width=\columnwidth]{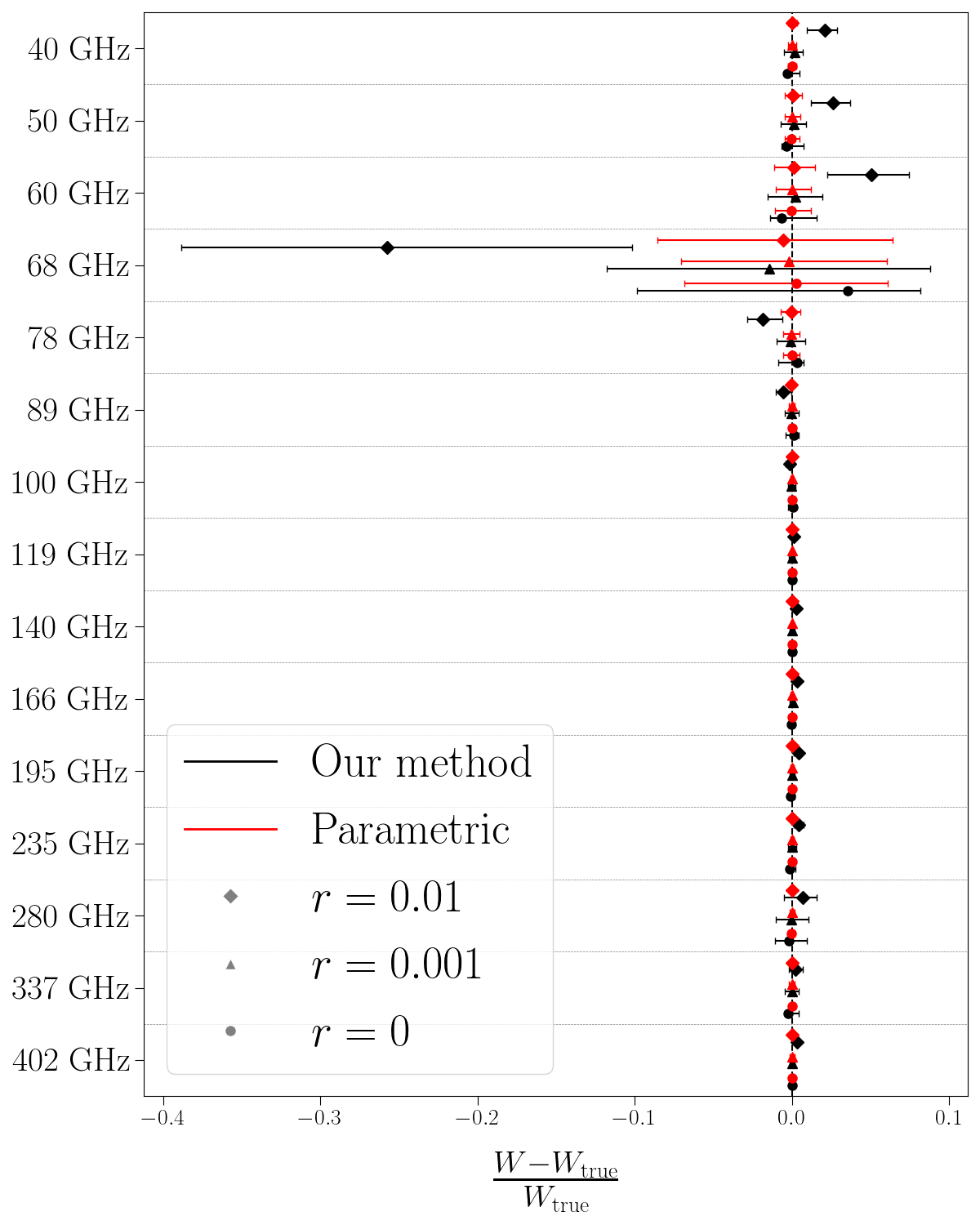}
\caption{Mean and 68\% confidence interval of the relative weights, compared with the true weights used as input on the simulations, for each of the frequency channels of the LiteBIRD-like case, for the three cases $r = 0, 0.001 \text{ and } 0.01$. In all cases, the model of foregrounds used as reference is \texttt{d0s0}.}
\label{fig:weights LB d0s0}
\end{center}
\end{figure}

Following the same steps as in the previous section, we show the weights of the frequency maps to produce the reconstructed CMB map for the two methods in FIG.~\ref{fig:weights LB d0s0}. We see that the weights are correctly reproduced by the two methods for $r = 0 \text{ and } 0.001$, but for $r=0.01$, the weights of the low frequency channels are biased in our method. However, it seems that overweighting the lowest frequencies and underweighting the intermediate frequencies can cancel out to recover an unbiased constraint on the tensor-to-scalar ratio. 

\subsection{Non-standard SED}

Now that we checked that our method performs properly in standard cases of interest, we can turn to what constitutes the real advantage of non-parametric methods. Indeed, parametric component separation methods can perform poorly when the foreground models do not accurately reproduce the spectral behaviour of foreground components \cite{Wolz:2023lzb,Errard:2022fcm,Remazeilles:2015hpa}. In particular, this would lead to a bias in the cosmological results from the mismodelling of foregrounds. In principle, because they are based on other assumptions, non-parametric methods should not be subject to this flaw and should be able to properly reconstruct the CMB signal whatever the actual spectral behaviour of the foreground components. To check whether this is indeed the case, we are now going to test the performance of both the parametric method and our approach on a sky where the dust component follows an unknown SED. For simplicity, we consider this case in the context of the SO-SAT-like experiments presented in Section \ref{section:Implementation of the spectral likelihood}. \\

\paragraph{Spatially homogeneous}
\ \\
In order to check the robustness of our method, we test it first on a spatially homogeneous dust model. Indeed, the method described in this paper is not suited to tackle spatial variations of the parameters. Because the built-in non-standard \texttt{PySM} models are either very close to a MBB spectrum or include some level of spatial variations, we use a modified version of the \texttt{d4} model of \texttt{PySM} \cite{Thorne:2016ifb,Zonca:2021row}. This model includes 2 populations of dust, each following a MBB SED, with specific values of the spectral parameter and temperature. Because the native \texttt{d4} model includes spatial variations of the SED through a temperature template based on \cite{Meisner-2015ApJ...798...88M}, we modified the \texttt{d4} model to remove the spatial variations of the dust temperatures by using their average on the sky. The first dust population has temperature $T_{d_{1}} = 9.73 \ \mathrm{K}$ and spectral index $\beta_{d_{1}} = 1.63$, while the second dust population has temperature  $T_{d_{2}} = 15.67 \ \mathrm{K}$ and spectral index $\beta_{d_{2}} = 2.82$. Secondly, the fact that there are two dust populations with different SEDs will also appear as spatial variations of the SEDs unless we assume the presence of an extra foreground component to take into account the two dust populations. In order to avoid this problem, we use the same template maps for the two dust populations. In other words, the SED of dust is now the sum of two MBBs, with homogeneous scaling across the sky. In the following, we refer to this model of dust as ``homogeneized \texttt{d4}''. This is a simplified case to test the properties of our method that is currently only suited for homogeneous scaling. Tackling spatial variations, and therefore handling more realistic sky signal. is left for future work.


\begin{figure*}[!htb]
\begin{center}
\includegraphics[width=1.5\columnwidth]{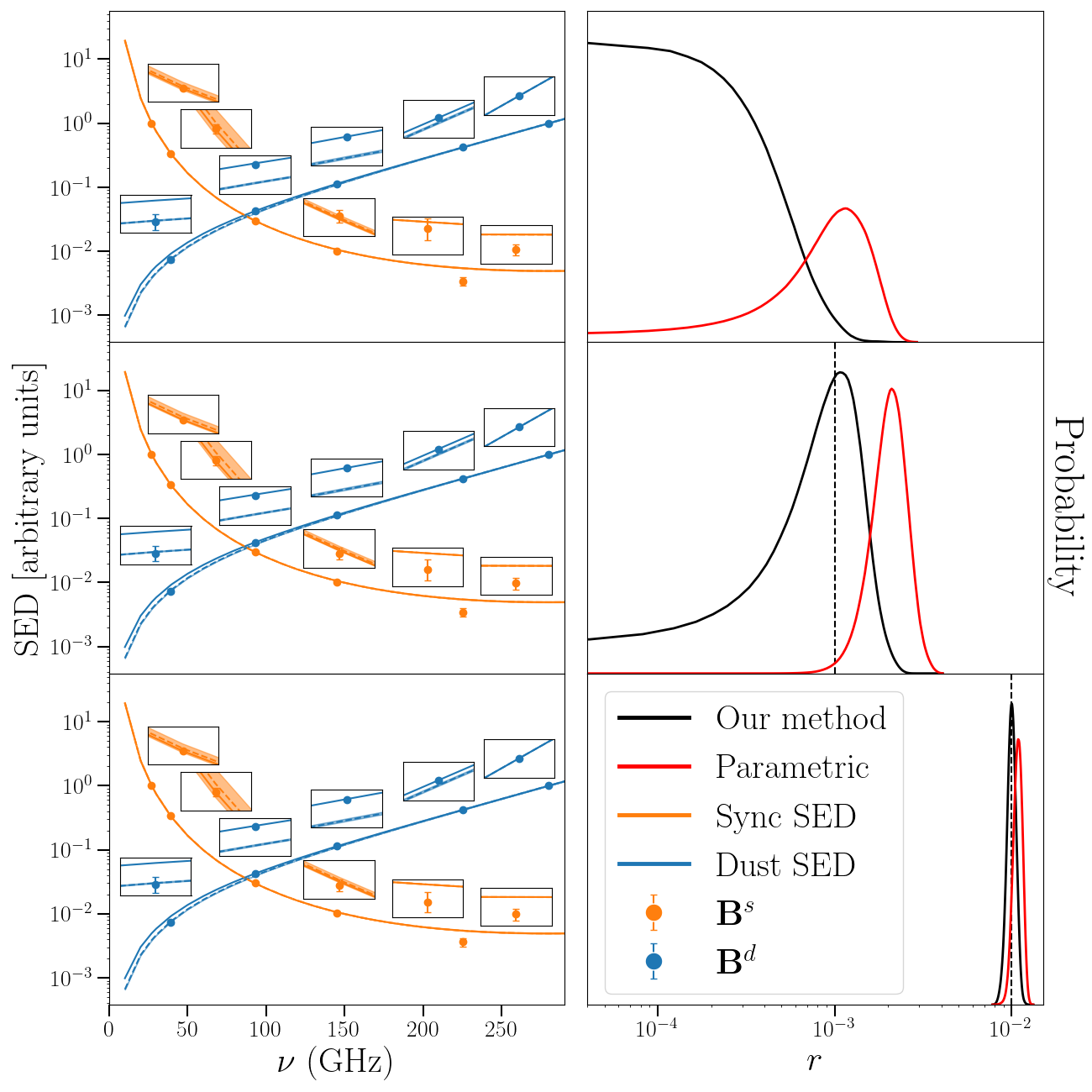}
\caption{Comparison of the constraints on the spectral parameters in the SO-SAT-like case when using the ``homogeneized \texttt{d4}'' dust component (Left) and tensor-to-scalar ratio (Right) when the true value of $r$ is 0.01 (Top), 0.001 (Middle) and 0 (Bottom) from the standard parametric approach of \texttt{FGBuster} and the method described in the present paper. The dots with error bars correspond to the mean and 68\% confidence interval of the marginal posterior of the mixing matrix elements from our method, the mean and 68\% confidence intervals of the SEDs from the parametric approach correspond to the dashed lines and shaded areas. The true SEDs are represented by the solid lines, in particular the model of synchrotron used is \texttt{s0}.}
\label{fig:results SO-SAT-like d4s0}
\end{center}
\end{figure*}

In the case of a sky signal with dust component following the ``homogeneized \texttt{d4}'' SED, we see in FIG.~\ref{fig:results SO-SAT-like d4s0} that the parametric method fails to reproduce the correct SED of the dust component, which in turn also leads to a degradation of the reconstruction of the synchrotron SED. This leads to a significant bias on the reconstructed value of the tensor-to-scalar ratio. On the contrary, our method which does not assume scaling laws for the foreground components is able to correctly constrain the elements of the mixing matrix to reproduce the foreground SEDs, in particular the SED of the dust component, which results in an unbiased measurement of $r$.

\begin{figure}[!htb]
\begin{center}
\includegraphics[width=\columnwidth]{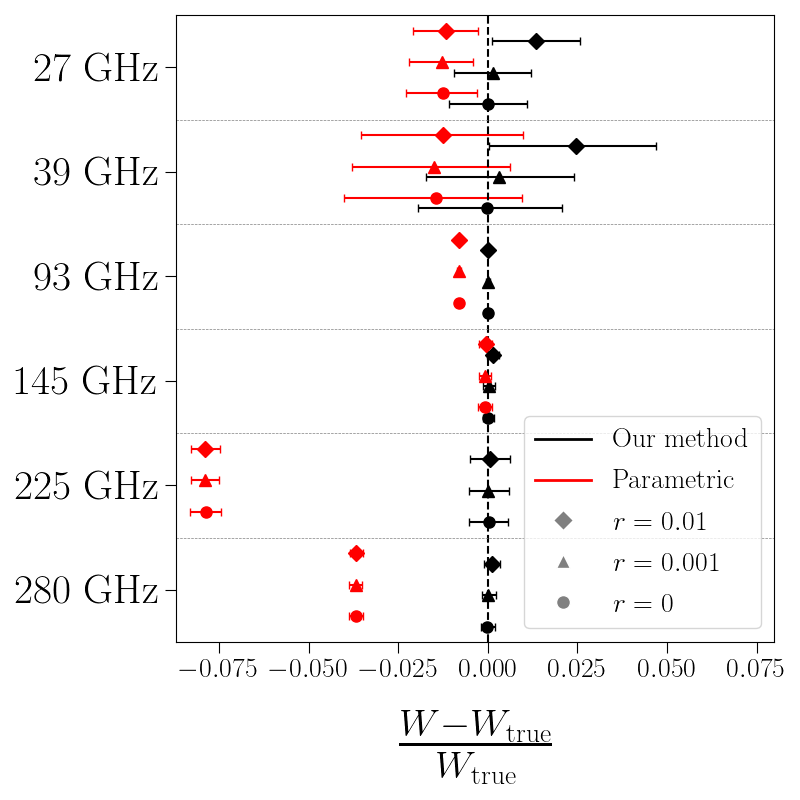}
\caption{Mean and 68\% confidence interval of the relative weights using the "homogeneized \texttt{d4}" dust component, compared with the true weights used as input on the simulations, for each of the frequency channels of the LiteBIRD-like case, for the three cases $r = 0, 0.001 \text{ and } 0.01$. In all cases, the model of synchrotron used is \texttt{s0}.}
\label{fig:weights SO d4s0}
\end{center}
\end{figure}

The weights applied to the frequency maps to reconstruct the CMB signal, whose comparison with their true values are presented in FIG.~\ref{fig:weights SO d4s0}, show a consistent picture. We do not see significant differences with the results on the weights from the case using the \texttt{d0} model in FIG.~\ref{fig:weights SO d0s0}. On the other hand, from the mismatch between the model used in the simulations and the model used in the parametric component separation, the parametric weights for the high frequency channels dominated by the dust component are completely off, and there is an offset even in the low frequency channels dominated by synchrotron emissions. \\

\begin{figure}[t]
\begin{center}
\includegraphics[width=\columnwidth]{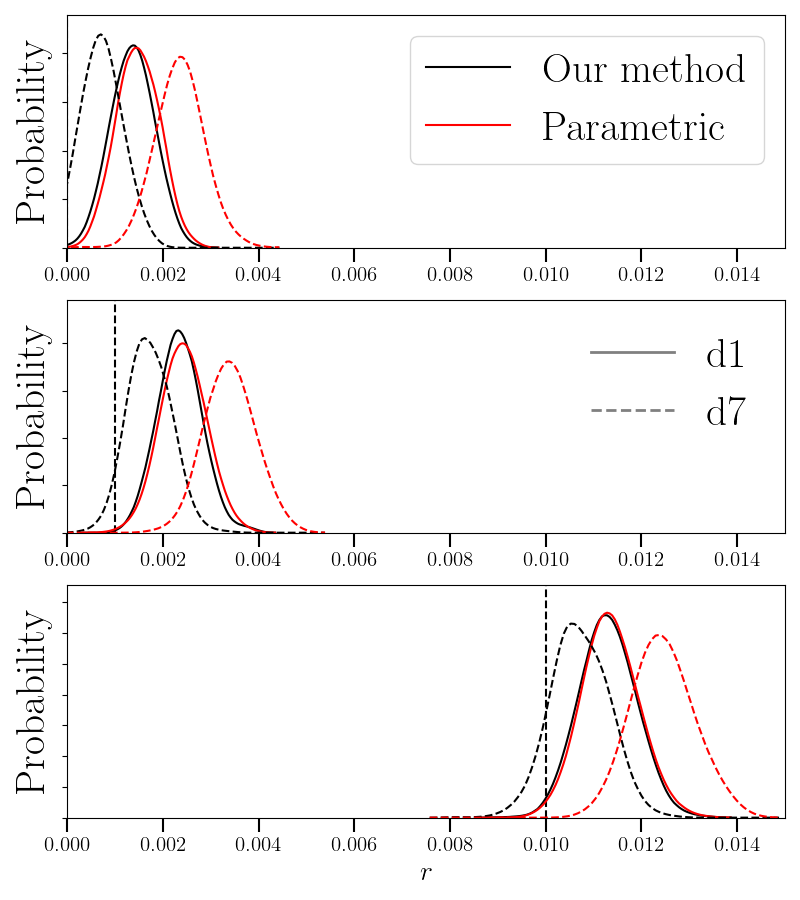}
\caption{Marginal posteriors of the tensor-to-scalar ratio $r$ for the two methods used in the present study when the Galactic dust follows the models \texttt{d1} (solid) and \texttt{d7} (dashed) of \texttt{PySM} and the true value of $r$ is 0 (Top), 0.001 (Middle) and 0.01 (Bottom). In all cases, the model of synchrotron used as reference is \texttt{s0}.}
\label{fig:results SO-SAT-like d1 and d7}
\end{center}
\end{figure}

\paragraph{Spatially varying}
\ \\
We turn now towards a dust component with a spatially varying SED, i.e. with a frequency scaling that depends on the sky pixel. We use the physically motivated built-in \texttt{PySM} models \texttt{d1}, which corresponds to a spatially varying MBB spectrum, and \texttt{d7}, based on the dust model of \cite{Hensley-d7} with additional iron inclusions in the grain composition \cite{Thorne:2016ifb,Zonca:2021row}. Because there is no single true SEDs to compare to, we describe the performances of the two methods from the measurement of the tensor-to-scalar ratio only. The posterior distributions of $r$ are given in FIG.~\ref{fig:results SO-SAT-like d1 and d7}. The presence of a significant bias in all studied cases is clearly due to the fact that the foreground cleaning methods presented here are not suited to handle any spatial variations of the foreground SEDs. However, we can also see that both methods performs similarly for the model \texttt{d1}, which follows a MBB SED and is therefore correctly reproduced by the parametric approach up to spatial variations, while our method performs significantly better, i.e. significantly reduces the bias, for the model \texttt{d7} because it provides a better fit to the averaged SED on the observed sky patch from its greater flexibility.

\section{Conclusions}
\label{section:Conclusions}

In this work, we first discussed the role of different assumptions on the component separation problem for CMB data in a unified maximum likelihood-based framework. In particular, we linked the different sets of assumptions used in component separation methods to their capacity to set constraints on the elements of the mixing matrix. For instance, we were able to show that we can always redefine the component separation problem such  that it is possible to use a template fitting approach, as long as the frequency scaling of CMB is known. We also argue that the maximum likelihood component separation problem can be fully solved only if the SEDs of all components, CMB and foregrounds, are known or can be correctly parametrized with fewer than $n_\nu-n_c$ parameters per sky component, or if a properly parametrized prior on the statistical properties is available for all the sky components. We have also shown that a more modest task of producing foreground-cleaned CMB maps can be achieved with significantly fewer assumptions.

Based on these results, we then proposed a minimally-informed, non-parametric component separation method for the measurement of primordial CMB $B$ modes. The method makes only the most basic assumptions about foreground components, such as their numbers and the spatial variability of their amplitudes, and relies instead on known properties of the CMB signal. We studied the method first analytically and later numerically. In particular, we developed a simple implementation of the method in the harmonic domain and used it to demonstrated its performance on a number of examples of current interest. We have found that this new approach for foreground cleaning is indeed very promising. It matches the performance of the parametric approach for the simplest parametric foreground models, and exceeds it for more complex models, when both methods are not allowed to account on spatial variability of the foreground properties. Further work is needed in order to understand the ability of the proposed method in the more complex cases. This will call for a fully-fledged, pixel-based, advanced implementation of the method. While this requires a significant implementation effort, techniques required for this already exist and are used, for instance, in the advanced implementations of the parametric method~\cite{Errard:2018ctl,Eriksen:2007mx}. This effort is already on-going, and will be addressed in future work.

\section*{Acknowledgements}

We would like to thank Davide Poletti for his constribution to early discussions about this project, and Dominic Beck for reading the manuscript. The authors acknowledge support of the French National Research Agency (Agence National de la Recherche) grants: ANR BxB and ANR B3DCMB. This work is also part of the SCIPOL project funded by the European Research Council (ERC) under the European Union’s Horizon 2020 research and innovation program (PI: Josquin Errard, Grant agreement No. 101044073). Some of the results in this paper have been derived using the \texttt{healpy} \cite{2005ApJ...622..759G,Zonca2019,healpix}, \texttt{numpy} and \texttt{PySM} packages. Some of the figures in this article have been created using \texttt{GetDist}.

\bibliography{bibliography}

\end{document}